\documentclass[%
reprint,
amsmath,amssymb,
prb,
floatfix,
longbibliography,
]{revtex4-2}

\usepackage[utf8]{inputenc}

\usepackage[top=56pt,bottom=56pt,left=50pt,right=50pt]{geometry} 

\usepackage{graphicx}% Include figure files
\usepackage{bm}% bold math

\usepackage{siunitx}

\usepackage{xcolor}
\definecolor{linkRed}{rgb}{0.735683, 0.215906, 0.330245}

\usepackage{hyperref}
\hypersetup{pdfpagemode={UseOutlines},
bookmarksopen=true,
bookmarksopenlevel=0,
colorlinks=true,% Set to false to disable coloring links
citecolor=linkRed,% The color of citations
linkcolor=linkRed,% The color of references to document elements (sections, figures, etc)
urlcolor=linkRed,% The color of hyperlinks (URLs)
pdfstartview={Fit},
unicode=true,
breaklinks=false,
}

\usepackage[version=4]{mhchem}

\DeclareSIUnit\angstrom{\text {Å}}  

\usepackage{float}

\begin{document}

\title{First-principles calculation of electron-phonon coupling in doped \ce{KTaO3}}
\author{Tobias Esswein}
\email{tobias.esswein@mat.ethz.ch}
\author{Nicola A. Spaldin}
\email{nicola.spaldin@mat.ethz.ch}
\affiliation{Materials Theory, Department of Materials, ETH Zurich, Switzerland}
\date{\today}

%-----------------------------------------------------------------------------
%-----------------------------------------------------------------------------
\begin{abstract}
Motivated by the recent experimental discovery of strongly surface-plane-dependent superconductivity at surfaces of \ce{KTaO3} single crystals, we calculate the electron-phonon coupling strength, $\lambda$, of doped \ce{KTaO3} along the reciprocal-space high-symmetry directions. 
Using the Wannier-function approach implemented in the EPW package, we calculate $\lambda$ across the experimentally covered doping range and compare its mode-resolved distribution along the [001], [110] and [111] reciprocal-space directions.
We find that the electron-phonon coupling is strongest in the optical modes around the $\Gamma$ point, with some distribution to higher $k$ values in the [001] direction.
The electron-phonon coupling strength as a function of doping has a dome-like shape in all three directions and its integrated total is largest in the [001] direction and smallest in the [111] direction, in contrast to the experimentally measured trends in critical temperatures.
This disagreement points to a non-BCS character of the superconductivity.
Instead, the strong localization of $\lambda$ in the soft optical modes around $\Gamma$ suggests an importance of ferroelectric soft-mode fluctuations, which is supported by our findings that the mode-resolved $\lambda$ values are strongly enhanced in polar structures.
The inclusion of spin-orbit coupling has negligible influence on our calculated mode-resolved $\lambda$ values.
\end{abstract}

%-----------------------------------------------------------------------------
%-----------------------------------------------------------------------------
\maketitle

%-----------------------------------------------------------------------------
%-----------------------------------------------------------------------------
\section{Introduction}\label{sec:intro}

Perovskite-structure potassium tantalate (\ce{KTaO3}, KTO) exhibits many interesting phenomena, resulting from its high dielectric constant \cite{BarrettDielectric1952}, strong spin orbit coupling \cite{UweRaman1980} and charged ionic layers \cite{StengelElectrostatic2011}. 
The strong spin-orbit coupling (SOC), caused mainly by the heavy tantalum ion, leads to a band splitting of up to \SI{400}{\milli eV} \cite{MattheissEnergy1972,UweRaman1980} and possible applications in spintronic devices \cite{NakamuraElectric2009,GuptaKTaO32022}. 
The high dielectric constant, associated with a quantum paraelectric state \cite{RowleyFerroelectric2014} similar to that of \ce{SrTiO3} (STO) \cite{MullerSrTiO31979}, indicates proximity to ferroelectricity, which is predicted to yield a large strain-dependent Rashba spin splitting \cite{TaoStraintunable2016,GastiasoroGeneralized2022}. 
The need to compensate the alternating charged ionic layers at the surfaces is predicted to induce lattice polarization in thin films \cite{GattinoniPrediction2022}, and leads to the accumulation of compensating charges at the surfaces of bulk samples \cite{StengelElectrostatic2011}. 
The origin and nature of the compensating charge are still open questions, with reports of conducting two-dimensional electron gases (2DEGs) \cite{KingSubband2012,Santander-SyroOrbital2012}, charge-density waves with strongly-localized electron polarons \cite{ReticcioliCompeting2022}, and terrace-like structures of alternating termination \cite{SetvinPolarity2018}, depending on the annealing atmosphere and temperature.

\begin{figure}[htb]
	\includegraphics[width=\linewidth,keepaspectratio]{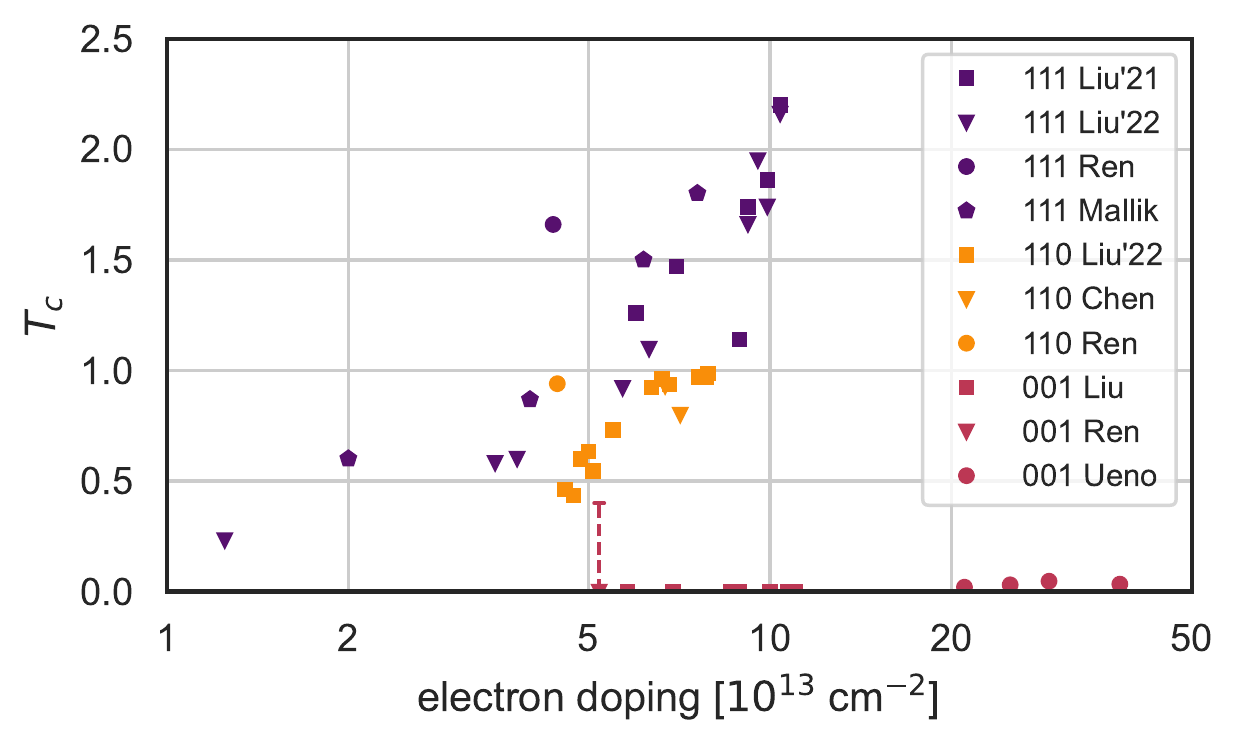}
	\caption{\label{fig:SC_Tc_exp}
        Superconducting critical temperatures, extracted from studies by 
        Ueno et al. \cite{UenoDiscovery2011}, 
        Chen et al. \cite{ChenTwoDimensional2021}, 
        Liu et al. \cite{LiuTwodimensional2021,LiuTunable2022}, 
        Ren et al. \cite{RenTwodimensional2022}, and 
        Mallik et al. \cite{MallikSuperfluid2022}.
		The (111) surface/interface reaches the highest T\textsubscript{c} of up to \SI{2}{K} (dark blue markers), followed by the (110) surface/interface reaching almost \SI{1}{K} (bright yellow markers).
		The original paper by Ueno et al. \cite{UenoDiscovery2011} reported a T\textsubscript{c} up to \SI{0.05}{K} for the (001) surface at high doping, but more recent publications at lower doping found no (001) superconductivity down to \SI{0.025}{K} \cite{LiuTwodimensional2021} and \SI{0.4}{K} \cite{RenTwodimensional2022} (red markers at bottom).
	}
\end{figure}

Perhaps the most intriguing behavior of KTO is its recently discovered low-temperature superconductivity on electron doping \cite{UenoDiscovery2011}. 
Superconductivity was first achieved using ionic liquid gating on the (001) surfaces of KTO single crystals, for which critical temperatures (T\textsubscript{c}) of up to \SI{50}{\milli K} were found at 2D doping concentrations of between \SI{2e14} and \SI{4e14}{cm^{-2}} \cite{UenoDiscovery2011, UenoFieldInduced2014}. Note that these values correspond to 3D doping concentrations of approximately \SIrange{4.1e20}{1.2e21}{cm^{-3}}, considerably higher than the $\sim$\SI{1.4e20}{cm^{-3}} possible using chemical doping with barium in bulk KTO \cite{SakaiThermoelectric2009}. 
(For the conversion between 2D and 3D carrier concentrations see Ref.~\onlinecite{UenoDiscovery2011} and the Appendix).
A subsequent study of LaAlO$_3$-capped KTO (110) surfaces, with 2D doping concentrations of \SI{7e13}{cm^{-2}}, reached markedly higher critical temperatures up to \SI{0.9}{K} \cite{ChenTwoDimensional2021}; (111)-oriented KTO interfaces with either \ce{EuO} or \ce{LaAlO3} showed even higher T\textsubscript{c}s of up to \SI{2.2}{K} at similar carrier concentrations \cite{LiuTwodimensional2021}. 
Note that no superconductivity was found down to \SI{25}{mK} at (001)-oriented KTO interfaces at these lower carrier concentrations \cite{LiuTwodimensional2021}. 
More recently, in an ionic liquid gating setup similar to that of  Ref.~\onlinecite{UenoDiscovery2011}, but at lower 2D doping densities of around \SI{5e13}{cm^{-2}}, superconductivity was found at the (110) and (111) surfaces with T\textsubscript{c} of around \SI{1}{K} and \SI{2}{K} respectively, and not at the (001) surface down to \SI{0.4}{K} \cite{RenTwodimensional2022}.
The reported critical temperatures from the literature are collected as a function of carrier concentration in figure~\ref{fig:SC_Tc_exp}.

The mechanism underlying the superconductivity, as well as its strong and unusual dependence on the orientation of the surface or interfacial plane, are not yet established. Indeed, even in the related quantum paraelectric STO, in which superconductivity was found more than half a century ago \cite{SchooleySuperconductivity1964,SchooleyDependence1965}, the pairing mechanism remains a subject of heated debate (for a recent review see Ref.~\onlinecite{GastiasoroSuperconductivity2020}). 
While the persistence to low carrier concentrations \cite{SchooleyDependence1965} and the anomalous isotope effect \cite{StuckyIsotope2016} challenge conventional BCS theories \cite{BardeenMicroscopic1957,BardeenTheory1957}, it is likely that electron-phonon coupling in some form, as well as proximity to ferroelectricity \cite{EdgeQuantum2015,RuhmanSuperconductivity2016,CoakPressure2019,vanderMarelPossible2019,GastiasoroAnisotropic2020a,GastiasoroTheory2022} play a role. 
Spin-orbit coupling has also been implicated \cite{KoziiOddParity2015,KanasugiSpinorbitcoupled2018,KanasugiMultiorbital2019a,KoziiSuperconductivity2019a}, and would be consistent with the observed higher critical temperatures in KTO, with its heavy tantalum ion, compared to STO \cite{GastiasoroAnisotropic2020a,GastiasoroTheory2022,YuTheory2022}.
The surface-plane dependence in KTO is captured by a model in which out-of-plane polar displacements of the Ta and O ions allow a linear coupling of the transverse optical (TO) phonon to the electrons in the t\textsubscript{2g} (d\textsubscript{xy}, d\textsubscript{yx} and d\textsubscript{zx}) orbitals; this coupling would otherwise go to zero as the phonon wavevector \textit{q} approached $\Gamma$ \cite{LiuTunable2022}.
The strong dependence of the  superconducting T$_c$ on surface orientation is then explained by different inter-orbital hopping of electrons between adjacent tantalum sites via the oxygen orbitals, with the highest hopping at (111) surfaces, followed by (110) surfaces, and no hopping allowed by symmetry at (001) surfaces.

It is clear that a thorough picture of the electron-phonon coupling as a function of electron doping and throughout the Brillouin Zone in KTO is an essential step towards developing a complete theory of its superconductivity. 
While the electron-phonon coupling has been calculated from first principles for STO \cite{ZhouElectronPhonon2018}, to our knowledge it is lacking for KTO, and the goal of this work is to remedy this gap.
Here we report the mode-resolved electron-phonon coupling strengths, $\lambda$, obtained using first-principles calculations based on density functional theory, for cubic \ce{KTaO3} across the range of experimentally accessible electron doping values.
We extract the mode-resolved total $\lambda$ as a function of carrier density, and focus in particular on differences between the [001], [110] and [111] high-symmetry directions, which are reciprocal to the corresponding experimentally measured surface and interfacial planes. 
Additionally, for one doping value, we compare the behavior with and without spin-orbit coupling, and for polar and non-polar structures to determine the effect of both properties.
Our main findings are that 
i) the calculated total electron-phonon coupling strengths do not follow the measured trends in superconducting T$_c$; 
ii) $\lambda$ is concentrated in the optical modes around $\Gamma$ and polar distortions increase $\lambda$ by a factor of approximately five, suggesting a mechanism involving the polar soft mode; 
iii) spin-orbit coupling has negligible influence on the calculated electron-phonon coupling.

%-----------------------------------------------------------------------------
%-----------------------------------------------------------------------------
\section{Methods}\label{sec:methods}

To calculate the forces and total energies we use density functional theory within the generalized gradient approximation (GGA) as implemented in the Quantum ESPRESSO 7.0 and 7.1 codes \cite{GiannozziQUANTUM2009,GiannozziAdvanced2017,GiannozziQuantum2020}.
We describe the exchange and correlation using the PBEsol functional \cite{PerdewRestoring2008}, and perform the core-valence separation with the ultrasoft GBRV \cite{GarrityPseudopotentials2014,GarrityGBRV2019} and pslibrary (to compare results with and without spin-orbit coupling) pseudopotentials \cite{DalCorsoPseudopotentials2014}.
We use a kinetic energy cutoff of \SI{60}{Ry} (\SI{816}{eV}) for the wavefunctions and a $24\times24\times24$~k-point mesh including $\Gamma$ for all unit cells.
Doping is achieved in the range from \SIrange{0.0001}{0.1}{} electrons/formula unit (e/fu) using the background-charge method with Gaussian smearing of \SI{1}{meV} width.
Total energies are converged to \SI{1}{\micro eV} (\SI{7.35e-8}{Ry}) and forces to \SI{0.1}{meV/\angstrom} (\SI{3.89e-6}{Ry/bohr}).

Both unit-cell size and shape, as well as internal coordinates, are fully relaxed, resulting in a non-polar cubic perovskite structure with a lattice constant of \SI{3.988}{\angstrom}, which is very close to the experimental one of \SI{3.989}{\angstrom} \cite{WempleTransport1965}.
Phonons are calculated on a $4\times4\times4$~q-point mesh, with convergence tests on $6\times6\times6$ and $8\times8\times8$~q-point meshes showing only minor quantitative differences (see appendix on page~\pageref{sec:appendix}).
The resulting phonon dispersion for very low doping, using the PBEsol-relaxed unit cell, corresponds well with the room-temperature phonon dispersion calculated recently using Quantum Self-Consistent Ab Initio Lattice Dynamics (QSCAILD), which is based on DFT and a self-consistent sampling method to capture both thermal and quantum fluctuations \cite{MeierFinite2022}. Note that, since our study is for doped KTO, we neglect the LO-TO splitting, assuming that it will be screened by the metallicity. To our knowledge, the evolution of the LO-TO splitting from the insulating to the metallic state as a function of doping in transition-metal oxides has not been determined, and this would be an important topic for future work.

The electron-phonon coupling properties are calculated using the EPW 5.4.1 and 5.5 codes \cite{GiustinoElectronphonon2007,PonceEPW2016}, which are included in the Quantum ESPRESSO package.
The relevant electronic bands in \ce{KTaO3} are the three Ta-$5d$ t\textsubscript{2g} bands, which are reproduced using maximally localized Wannier functions as implemented in the Wannier90 code \cite{PizziWannier902020}, used internally by EPW.
The electron-phonon matrix elements are first calculated on coarse $24\times24\times24$~k-point and $4\times4\times4$~q-point meshes and then interpolated onto fine grids using maximally localized Wannier functions.
We use a random fine mesh with \si{1'000'000}~k points to calculate the mode-resolved electron-phonon coupling strengths, $\lambda_{q\nu}$, along a path between cubic high-symmetry points with 200 q~points between each point.
Convergence test results can be found in the appendix on page \pageref{sec:appendix}.

%-----------------------------------------------------------------------------
%-----------------------------------------------------------------------------
\section{Results and Discussion}\label{sec:results}

\begin{figure*}[htp]
	\includegraphics[width=.47\linewidth,keepaspectratio]{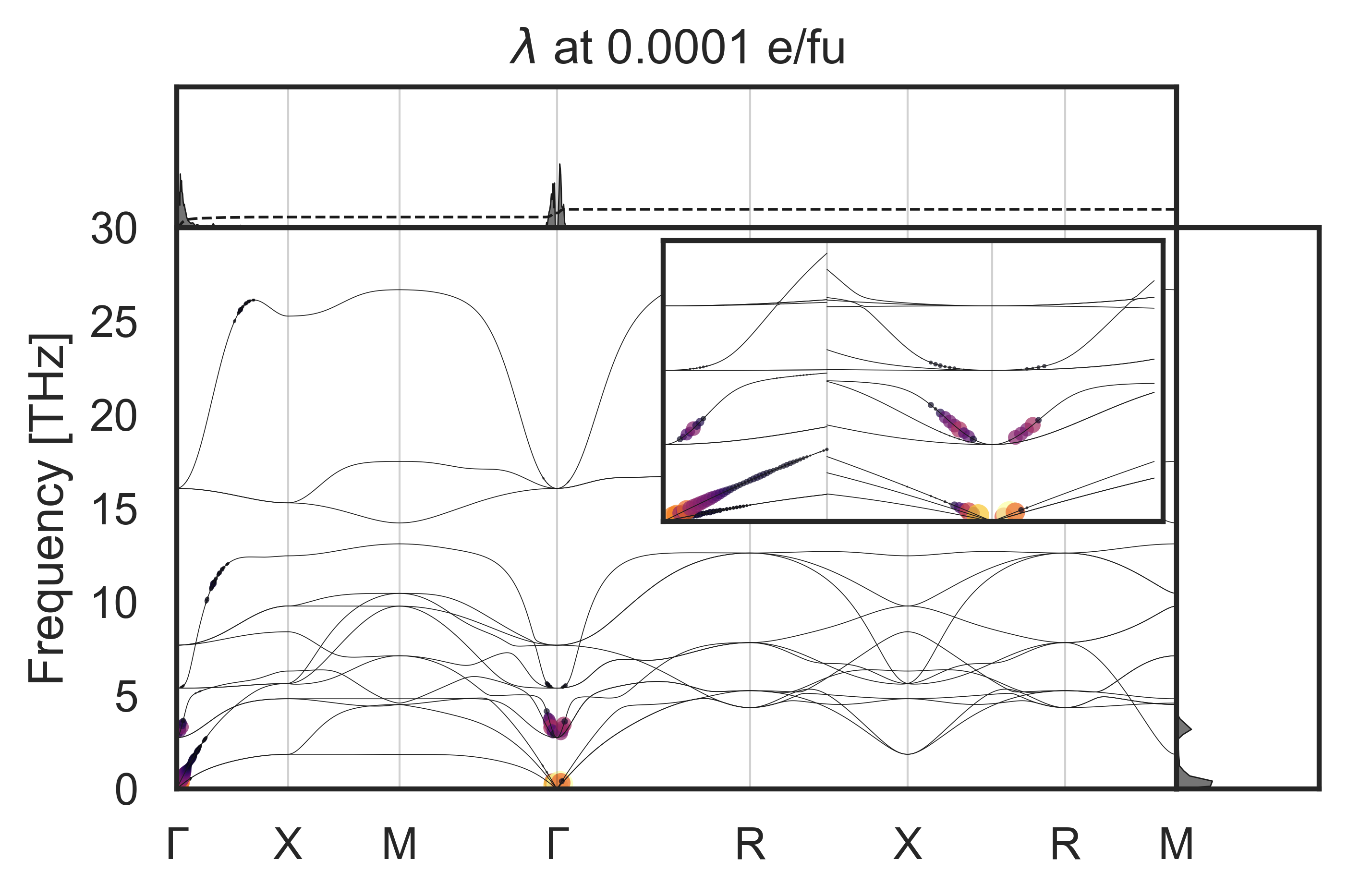}%
	\includegraphics[width=.47\linewidth,keepaspectratio]{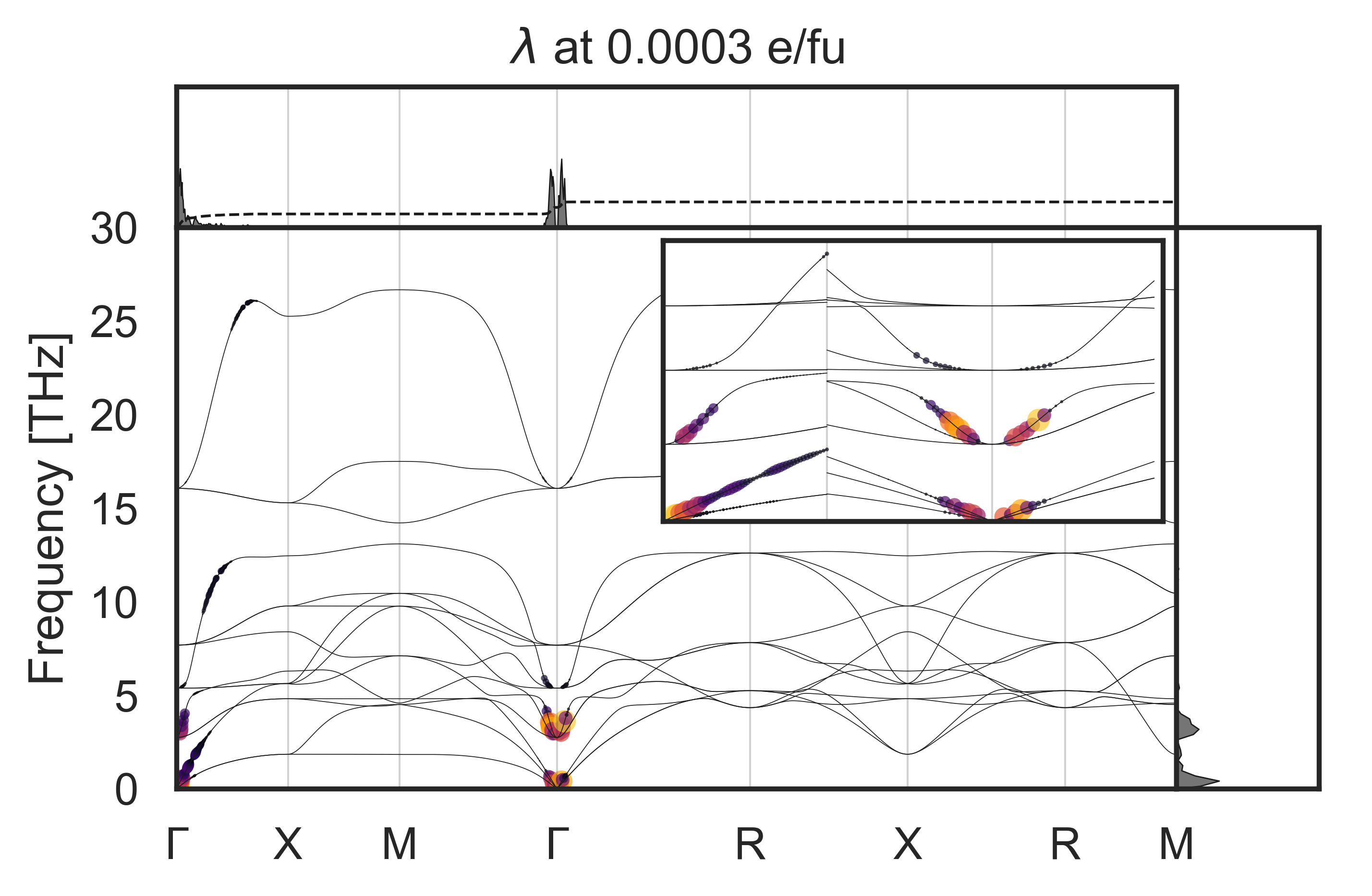}
	\includegraphics[width=.47\linewidth,keepaspectratio]{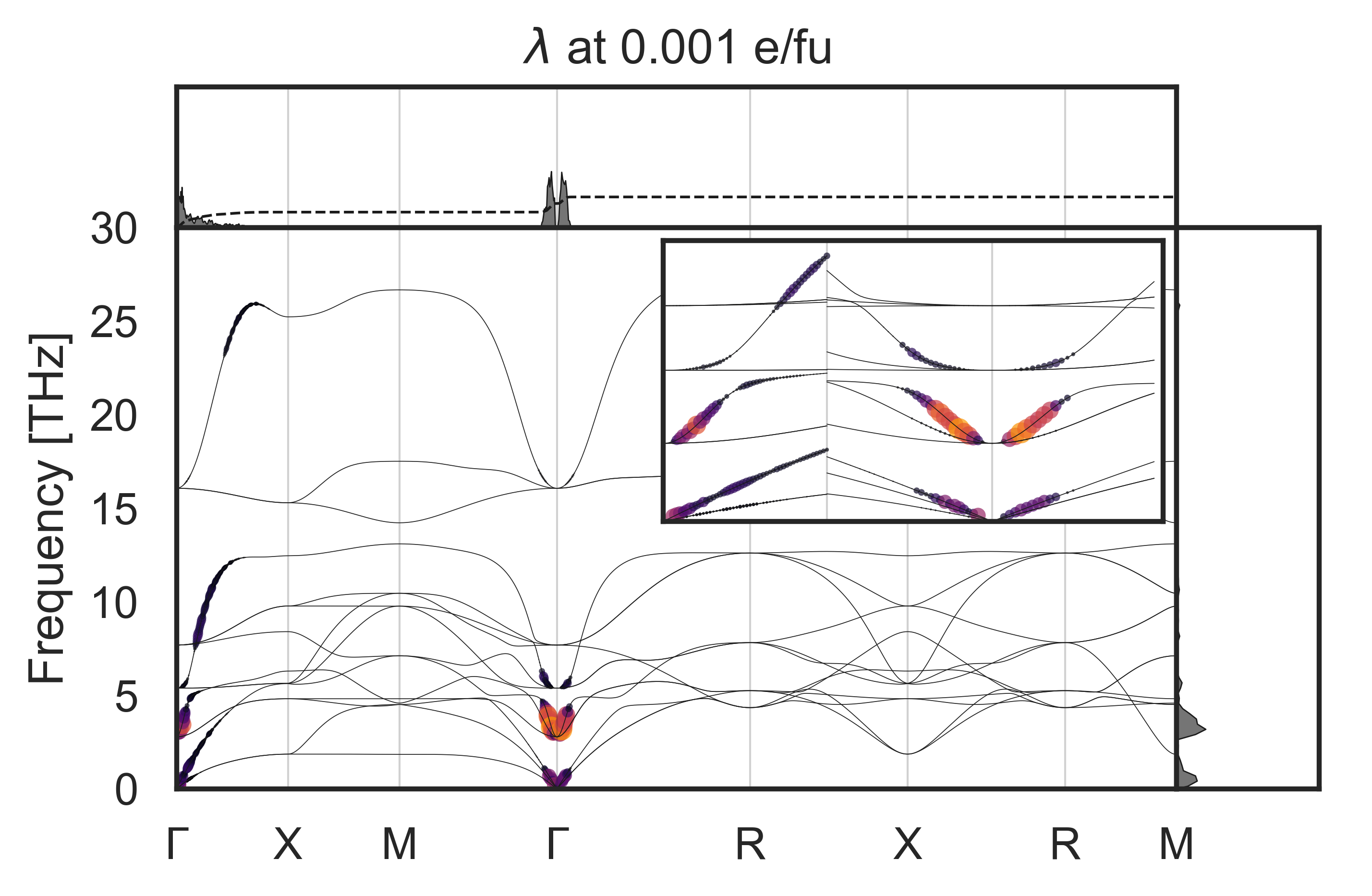}%
	\includegraphics[width=.47\linewidth,keepaspectratio]{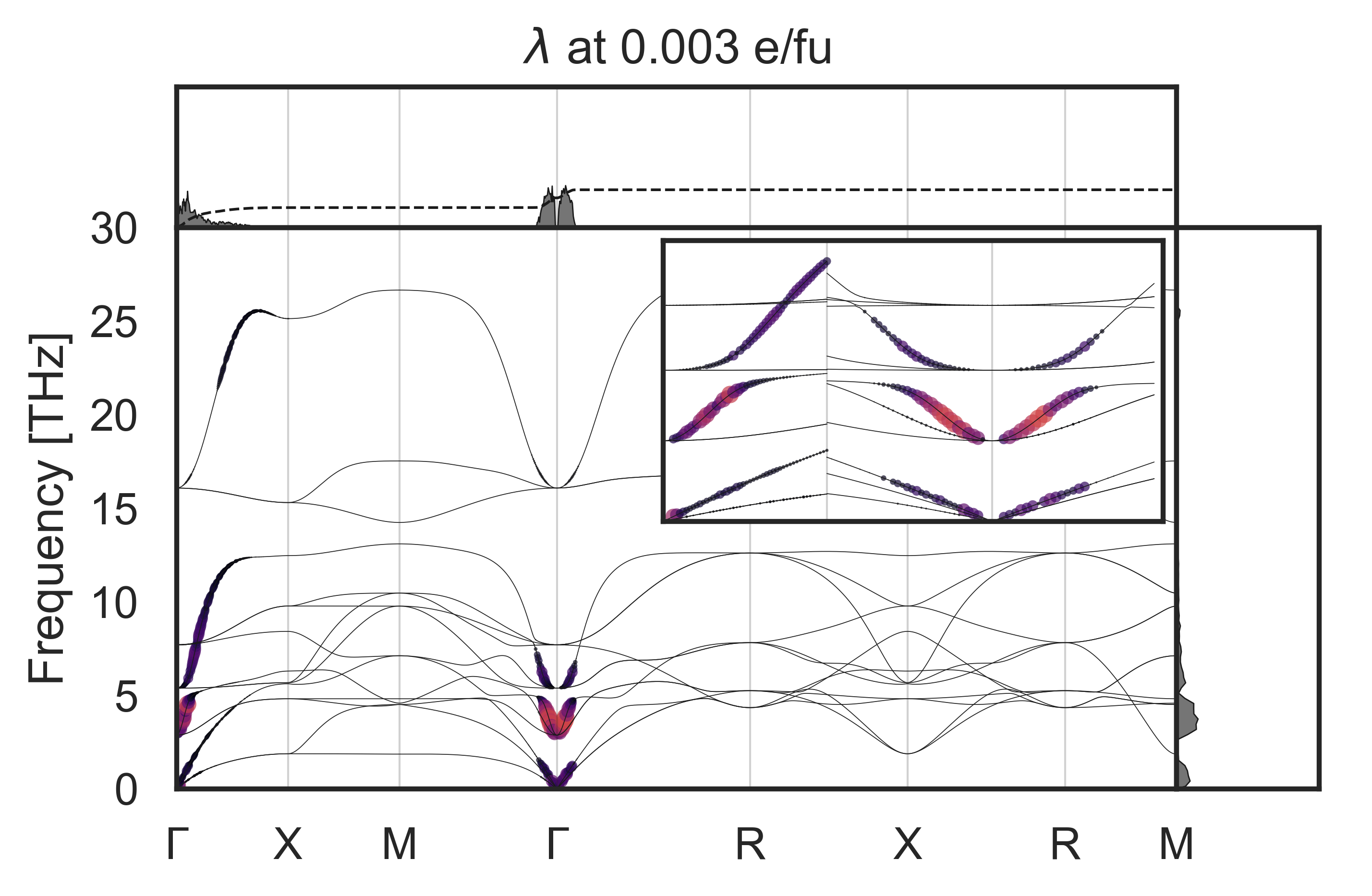}
	\includegraphics[width=.47\linewidth,keepaspectratio]{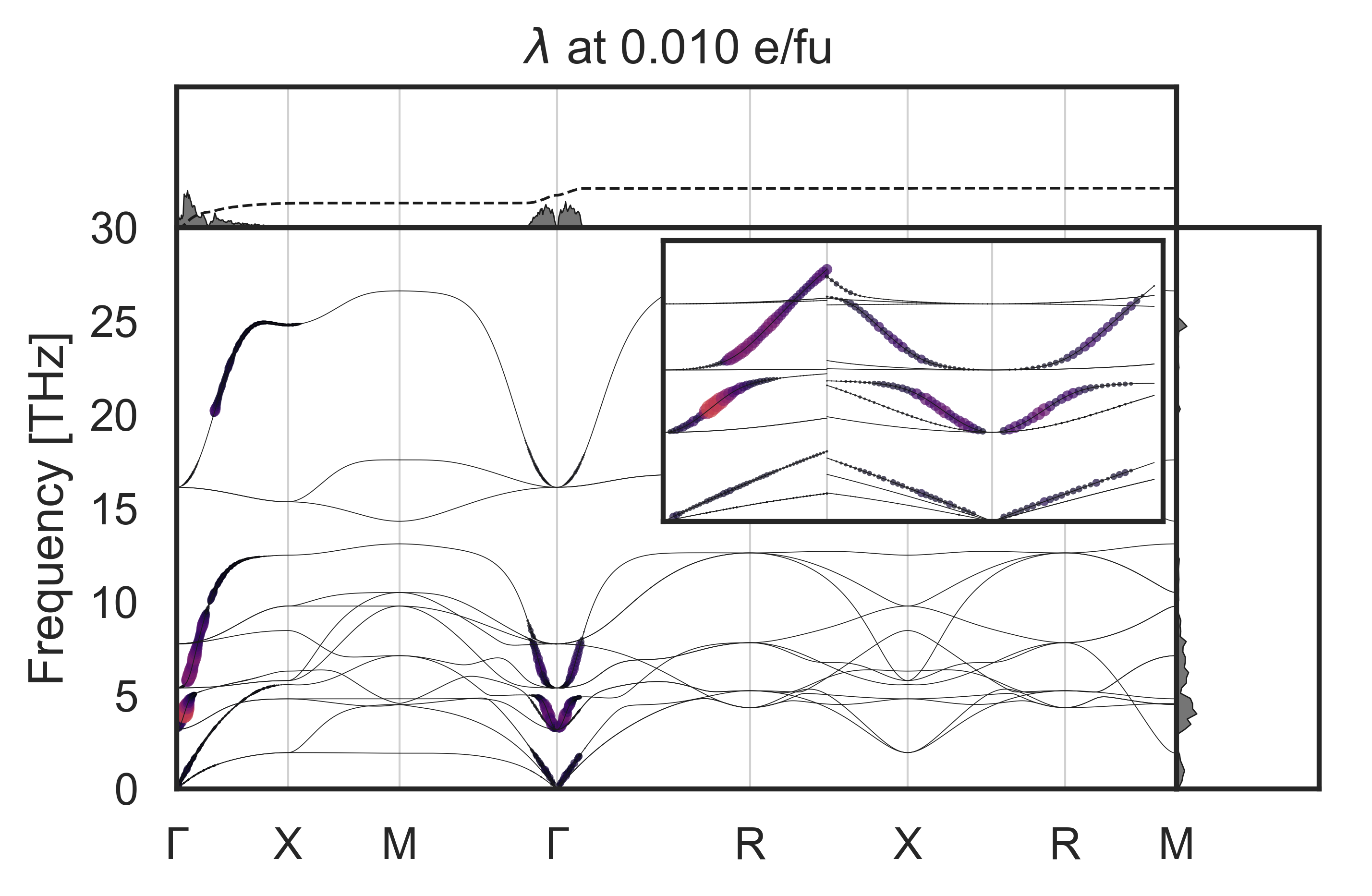}%
	\includegraphics[width=.47\linewidth,keepaspectratio]{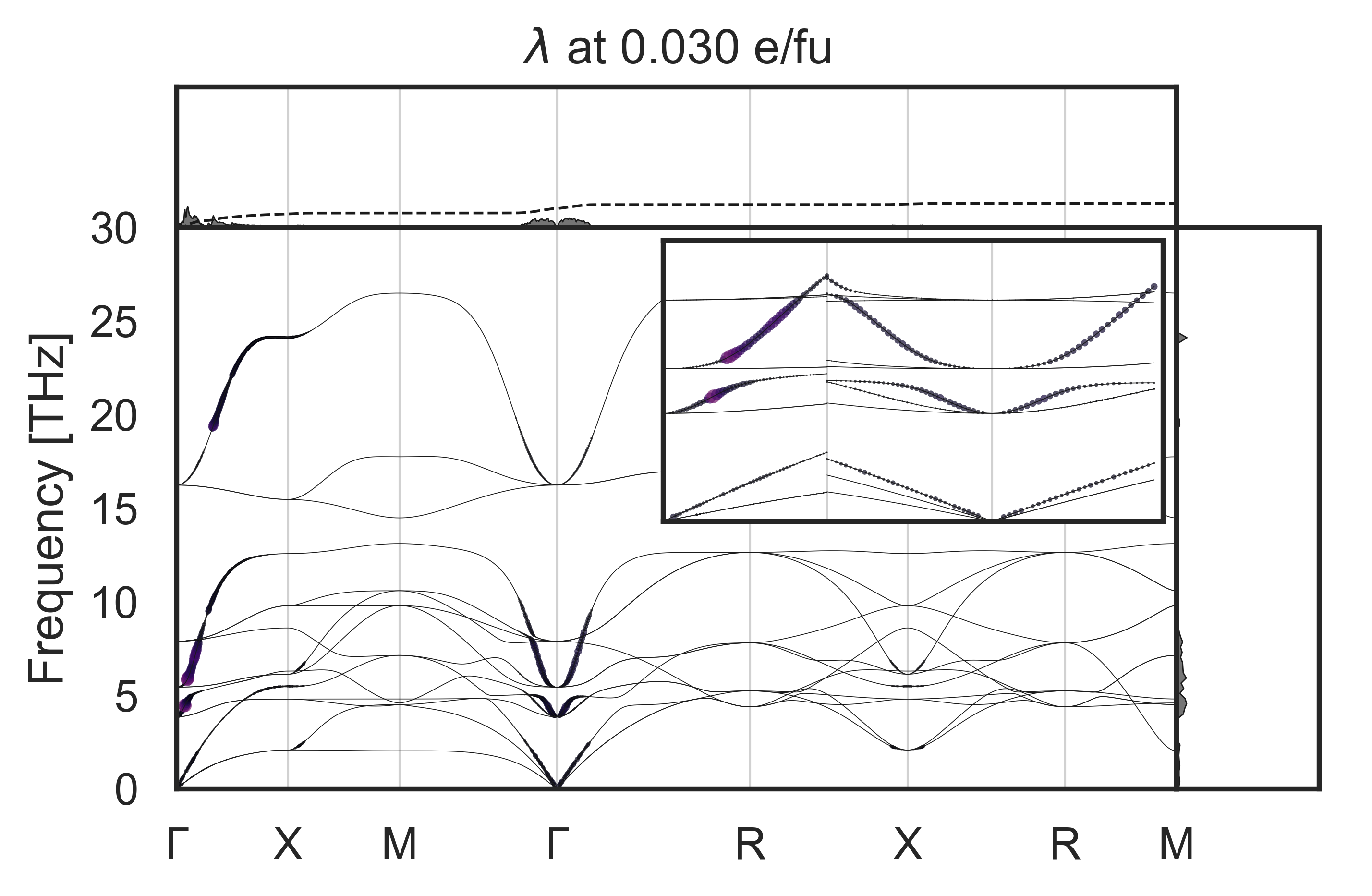}
	\includegraphics[width=.47\linewidth,keepaspectratio]{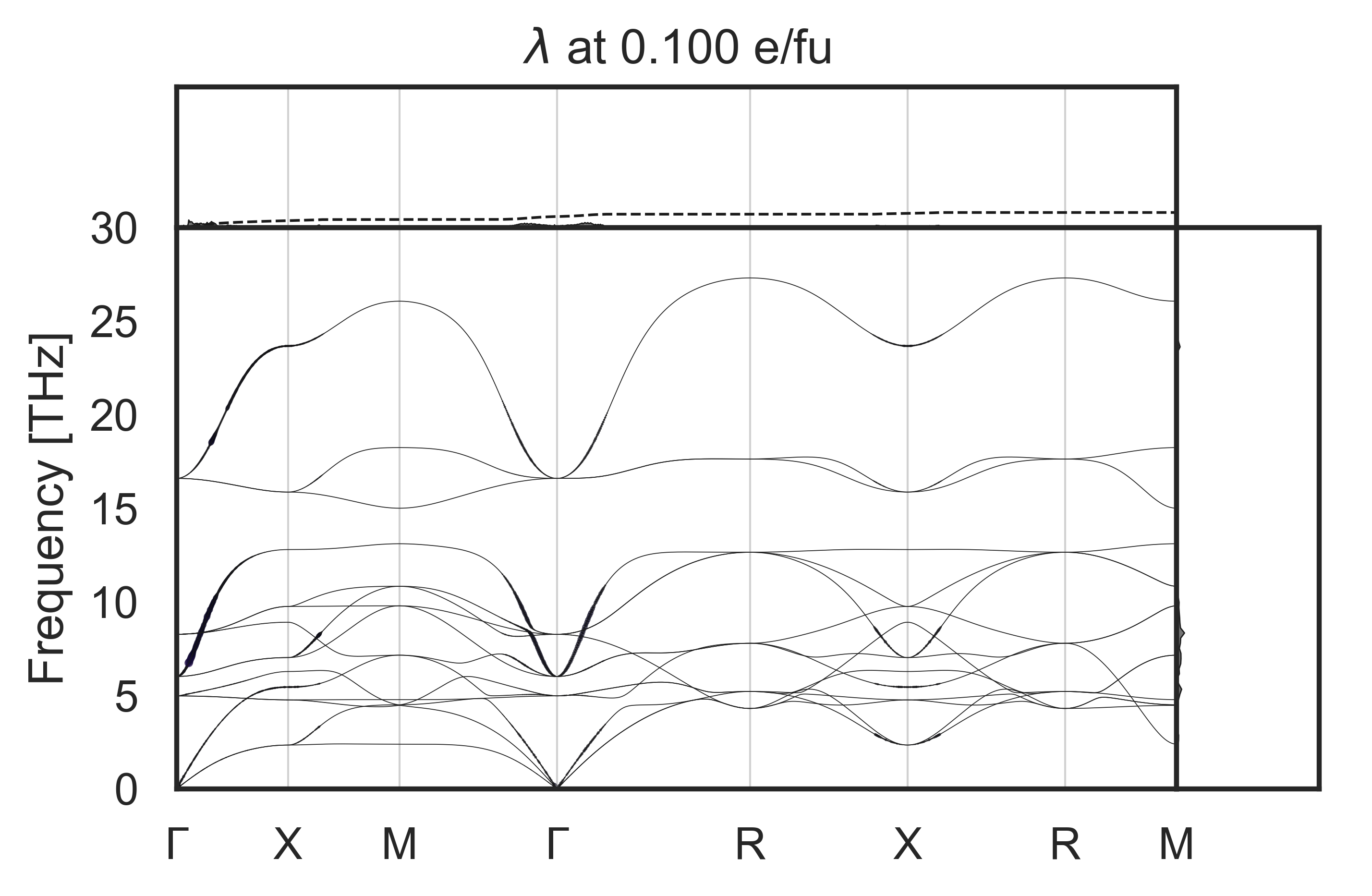}%
	\includegraphics[height=39mm,keepaspectratio,trim=0 -8mm 0 0,clip]{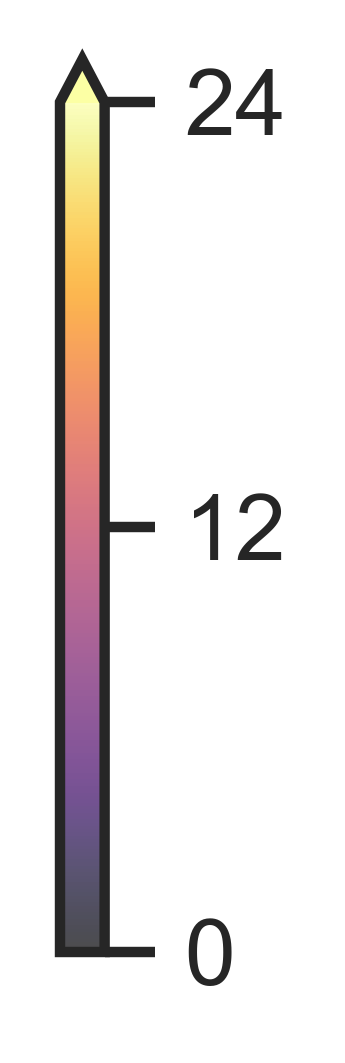}
	\caption{\label{fig:lambda_mode_resolved}
		Phonon-mode-resolved electron-phonon coupling strength $\lambda$ at different doping values ranging from \SIrange{0.0001}{0.1}{e/fu}, which correspond to \SIrange{1.6e+18}{1.6e+21}{e/cm^3} or roughly \SIrange{6.2e+12}{4.7e+14}{e/cm^2}.
		All plots are on the same scale and share the same colorbar for the dimensionless $\lambda$, shown at the bottom. 
        The insets in the top right corners of each subplot show a zoomed-in part of the area around $\Gamma$ towards the X, M and R points, ranging from from \SIrange{0}{10}{THz}.
        Integrated $\lambda$ values along vertical and horizontal directions are shown in the top and right subpanels, respectively.
	}
\end{figure*}

Our calculated mode-resolved electron-phonon coupling strengths $\lambda$ at seven different doping levels, covering the experimental range, are shown in figure~\ref{fig:lambda_mode_resolved}, along the high symmetry directions of the Brillouin zone, with $\lambda$ integrated at each q point shown in the top part of each subplot, and $\lambda$ integrated over frequency (decomposed into 100 frequency bins) shown on the right of each subplot.

There are several points to note.
First, there are no imaginary frequencies, as the structural relaxation of KTO using the PBEsol functional results in a cubic unit cell with no structural instabilities.
The frequency of the polar soft mode at $\Gamma$, which can be imaginary using the PBE functional, is \SI{2.7}{THz} for the lowest doping value of \SI{0.0001}{e/fu}, and hardens to \SI{5.0}{THz} at the highest doping value of \SI{0.1}{e/fu}. 
It has the strongest electron-phonon coupling strength $\lambda$ throughout the whole doping range.
Additionally, contributions to  $\lambda$ can be seen in the higher-energy optical modes around $\Gamma$. 
The strong coupling of the electrons to the polar modes at the $\Gamma$ point suggests that the ferroelectric fluctuations associated with quantum paraelectricity could play a key role in the superconductivity in KTO, as already suggested for quantum paraelectric STO \cite{EdgeQuantum2015,vanderMarelPossible2019,GastiasoroSuperconductivity2020}. 

In the [110] ($\Gamma$ to M) and [111] ($\Gamma$ to R) directions, the electron-phonon coupling occurs only close to the $\Gamma$ point; here the optical phonons correspond to long-wavelength ferroelectric displacements. 
In the [001] $\Gamma$-X direction, in contrast, the coupling, while strongest close to $\Gamma$, remains present along the entire high-symmetry line, also at higher doping. This results also in a larger total contribution along the [001] direction than along [110] and [111]. 
We note that the form of $\lambda$ in reciprocal space closely follows that of the Fermi surface, which at these doping levels is close to spherical except for elongations along the cartesian reciprocal axes reflecting the flat electronic bands along $\Gamma$ to X (see e.g. fig 3.3 of Ref.~\onlinecite{EssweinExploring2018}). 
As expected, at low doping, the electron-phonon coupling is limited largely to the lowest phonon frequencies, then extends to higher frequencies as the doping is increased and higher energy electronic bands are populated. 

\begin{figure}[ht]
	\includegraphics[width=\linewidth,keepaspectratio]{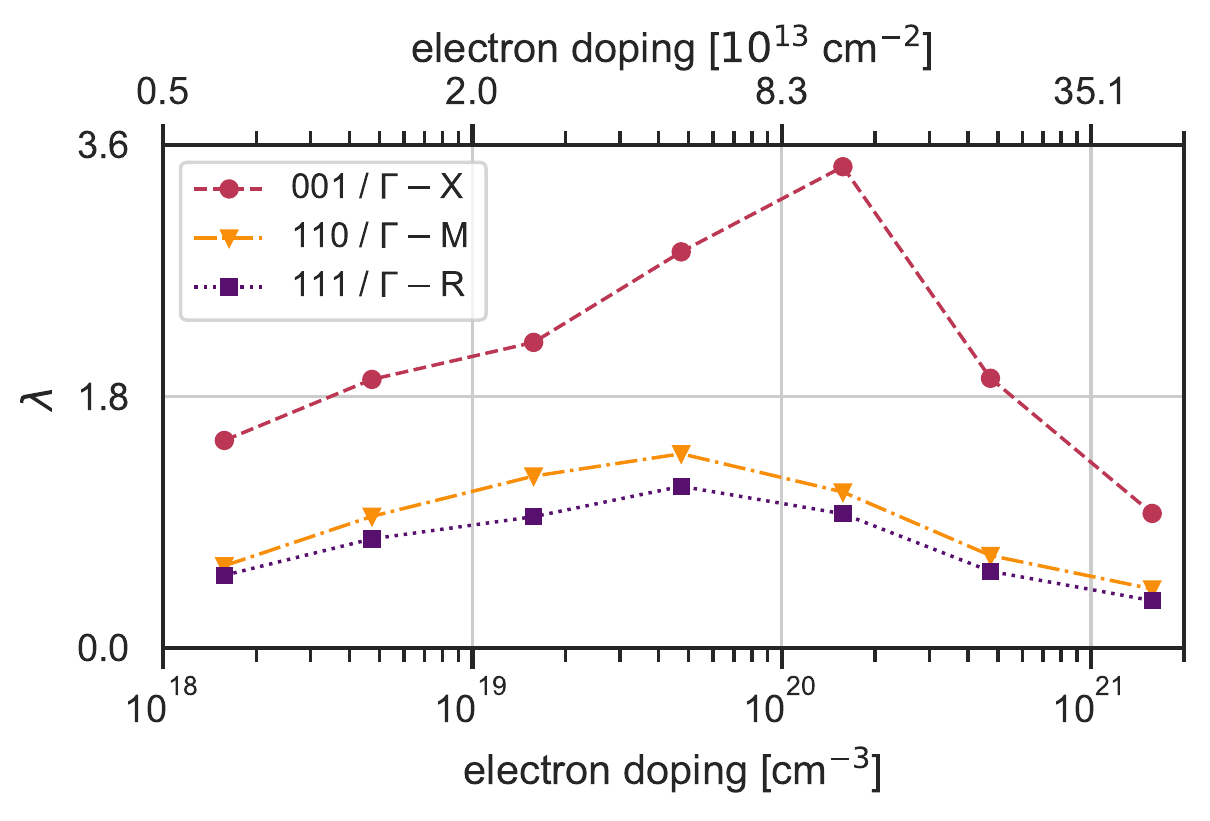}
	\caption{\label{fig:lambda_tot}
		Total electron-phonon coupling strength $\lambda$ integrated along each high-symmetry direction at different doping values, covering a range from low concentration to the maximum achieved by ionic liquid gating \cite{UenoDiscovery2011}.
		The numbers on the left axis correspond to the mean lambda value of each point along each high-symmetry direction.
		The 2D doping values on top are estimated from the 3D values on the bottom axis using the conversion method described in the appendix (see fig.~\ref{fig:interpol2D3D}).
		The strongest electron-phonon coupling is along the [001] direction, while the [110] and [111] directions have almost the same magnitude and evolution with doping.
	}
\end{figure}

The calculated integrated $\lambda$ values along the three high-symmetry directions, which we use as proxies for the total electron-phonon coupling strength in each reciprocal direction, are shown as a function of doping concentration in figure~\ref{fig:lambda_tot}. 
In all directions in reciprocal space there is a dome-like structure in the calculated $\lambda$, with a smooth maximum between \SIrange{1e19}{1e20}{e/cm^{-3}} (\SIrange{2.0e+13}{8.3e+13}{e/cm^{-2}}) for the [110] and [111] directions. 
The more pronounced peak around \SI{1e20}{e/cm^{-3}} in the [001] direction coincides with the electron doping reaching the X point of the phonon band structure, as can also be seen in the third row of figure~\ref{fig:lambda_mode_resolved}.

If KTO were a conventional BCS-theory superconductor, we would expect the critical temperatures of figure~\ref{fig:SC_Tc_exp} to follow roughly the electron-phonon coupling strength of figure~\ref{fig:lambda_tot}.
Comparing those two figures, it is clear that there is no obvious correlation. 
First, the experimental data do not show such a dome-like trend, with the (111) surfaces/interfaces in particular showing a linear increase of T\textsubscript{c} with increasing doping. 
Second, while experimentally the highest T\textsubscript{c} is observed for the (111) surfaces/interfaces, and the T\textsubscript{c} for the (001) surfaces/interfaces is very low, the electron-phonon coupling is strongest for the [001] direction, and weakest for the [111] direction.

%-----------------------------------------------------------------------------
\subsection*{Influence of spin-orbit coupling and polar symmetry breaking}
\begin{figure*}[htp]
	\includegraphics[width=.47\linewidth,keepaspectratio]{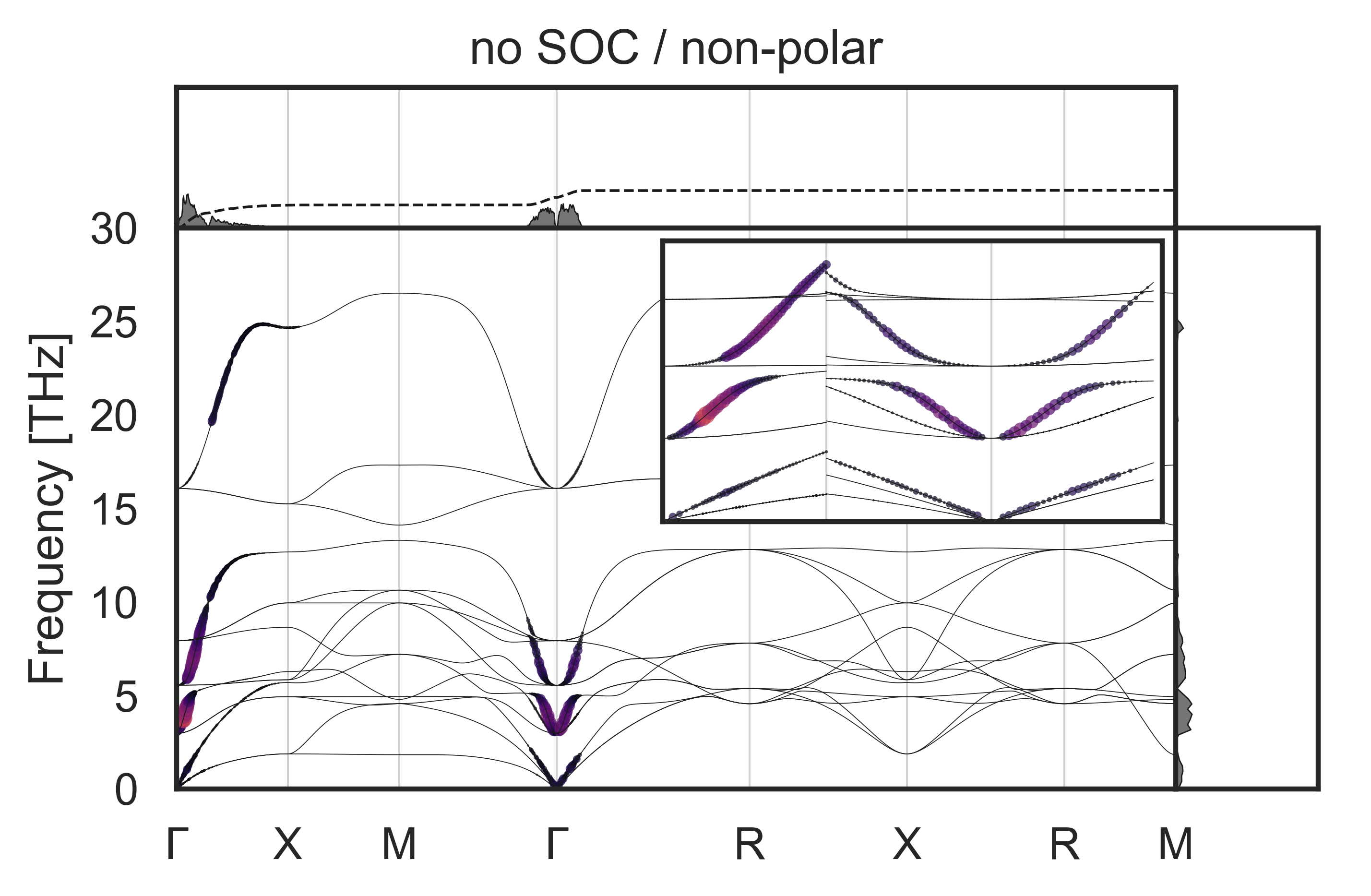}%
	\includegraphics[width=.47\linewidth,keepaspectratio]{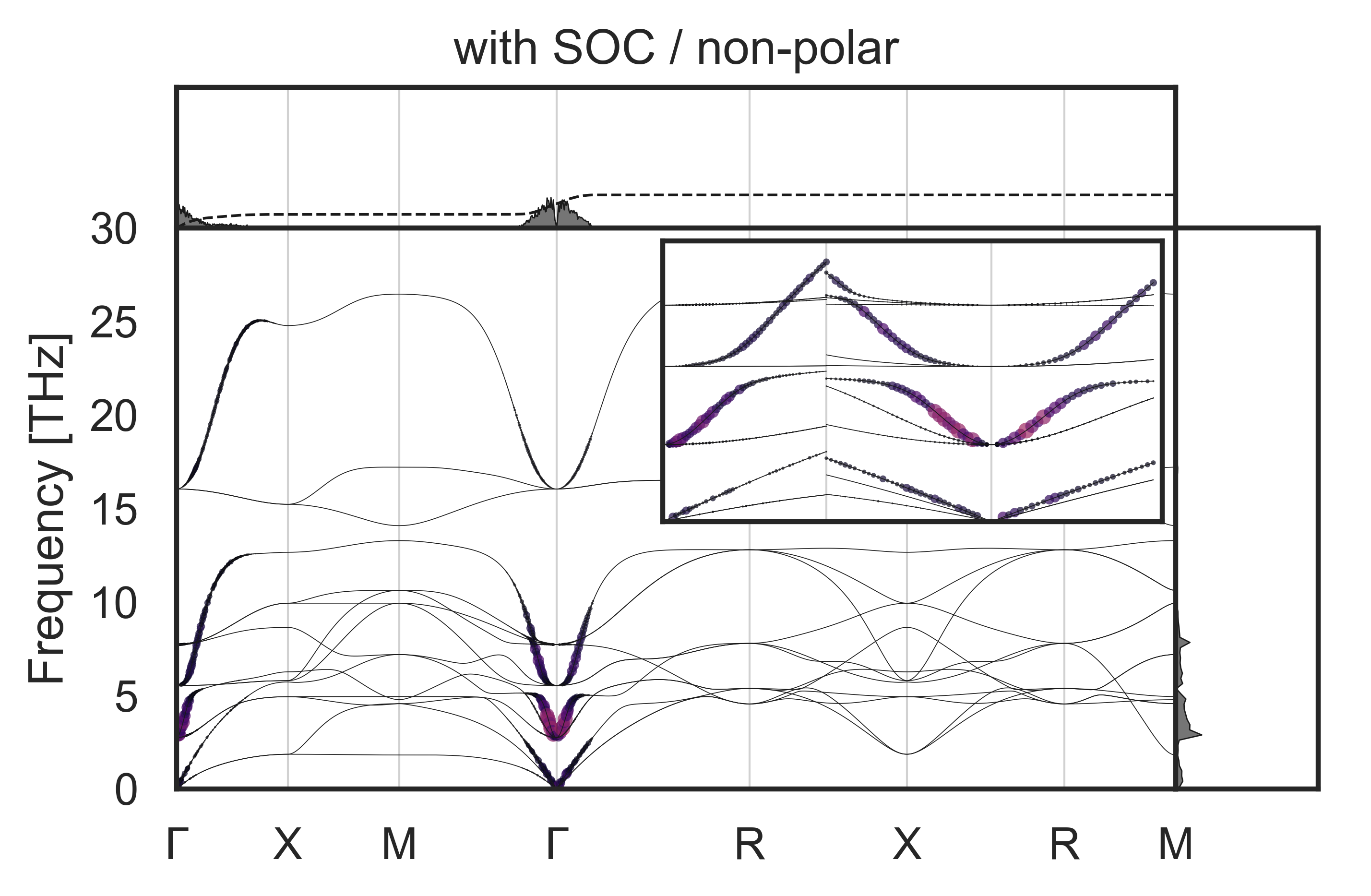}
	\includegraphics[width=.47\linewidth,keepaspectratio]{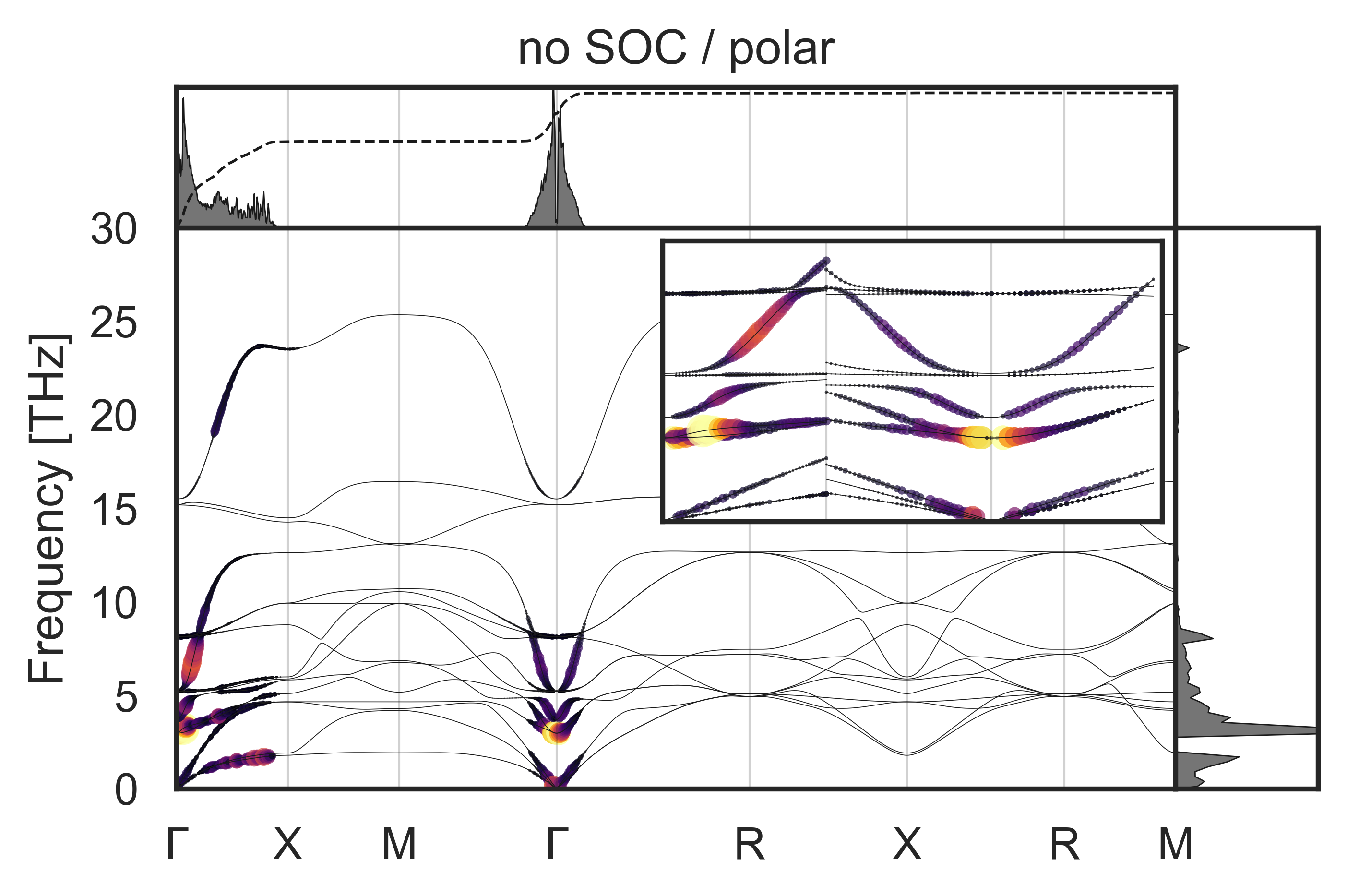}%
	\includegraphics[width=.47\linewidth,keepaspectratio]{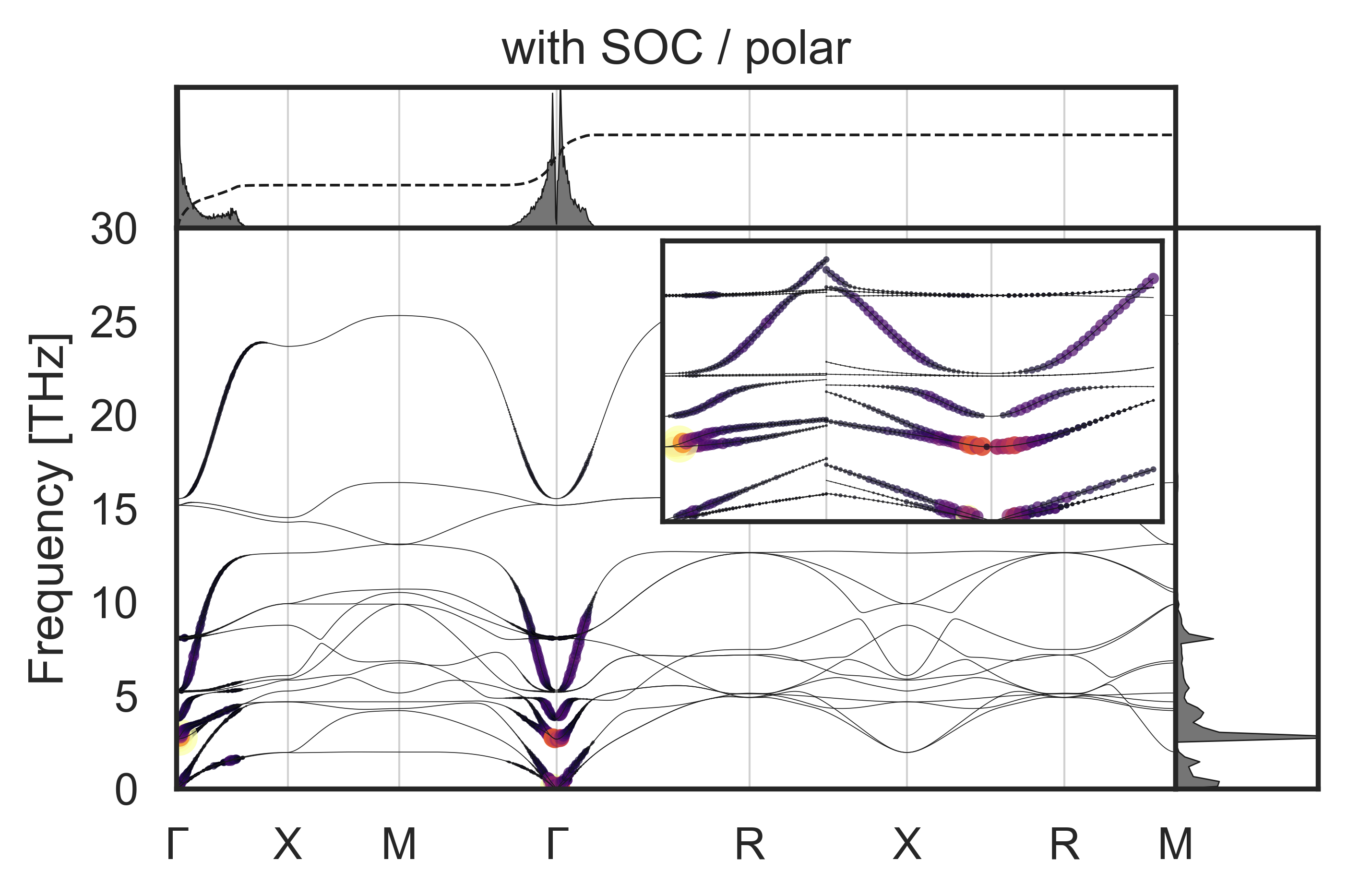}
	\caption{\label{fig:lambda_polar_soc}
		Calculated phonon-mode-resolved electron-phonon coupling strength $\lambda$ at a doping value of \SI{0.01}{e/fu} for non-polar (top) and polar (bottom) KTO without (left) and with (right) spin-orbit coupling (SOC).
		All plots share the same colorbar, shown at the bottom of figure~\ref{fig:lambda_mode_resolved}, for the dimensionless $\lambda$.
	}
\end{figure*}

Finally, to determine the influence of spin-orbit coupling and polar structural distortions, we present in figure~\ref{fig:lambda_polar_soc} our calculated mode-resolved $\lambda$ for both non-polar and [111]-polarized KTO with and without spin-orbit coupling (SOC).
(The influence of SOC on the electronic bands is shown in figure~\ref{fig:bands_SOC}.)
The polar structure is obtained by increasing the cubic lattice constant to \SI{4.010}{\angstrom} and relaxing the internal coordinates; the soft mode frequency is then close to that of the original cubic structure ($\sim$\SI{3.0}{THz}).

First we compare the results with and without SOC in the cubic structures (top row) \footnote{Note the small difference between the `no SOC/non-polar' panel of fig.~\ref{fig:lambda_polar_soc} and the corresponding `$\lambda$ at 0.010~e/fu' panel of fig.~\ref{fig:lambda_mode_resolved}.
The difference arises from the fact that for fig.~\ref{fig:lambda_polar_soc} we calculated everything using scalar-relativistic pslibrary pseudopotentials to make it directly comparable to the SOC results which we obtained using fully-relativistic pslibrary pseudopotentials.} and see that the differences are negligible.
In contrast, the differences between the non-polar and polar structures (down the columns) are substantial.
First, the integrated values of $\lambda$ in the polar structures are larger by a factor of around five than the non-polar values.
Second, the lowest energy branches of the soft TO modes around $\Gamma$, which had negligible electron-phonon coupling in the cubic structures, are now among the main contributors to $\lambda$.
Third, coupling along the $\Gamma$-X direction is enhanced by the polar distortion, especially in the polar case without SOC.
Note that in the polar case, spin-orbit coupling slightly reduces the $\lambda$ values.

\begin{figure}[!h]
	\includegraphics[width=\linewidth,keepaspectratio]{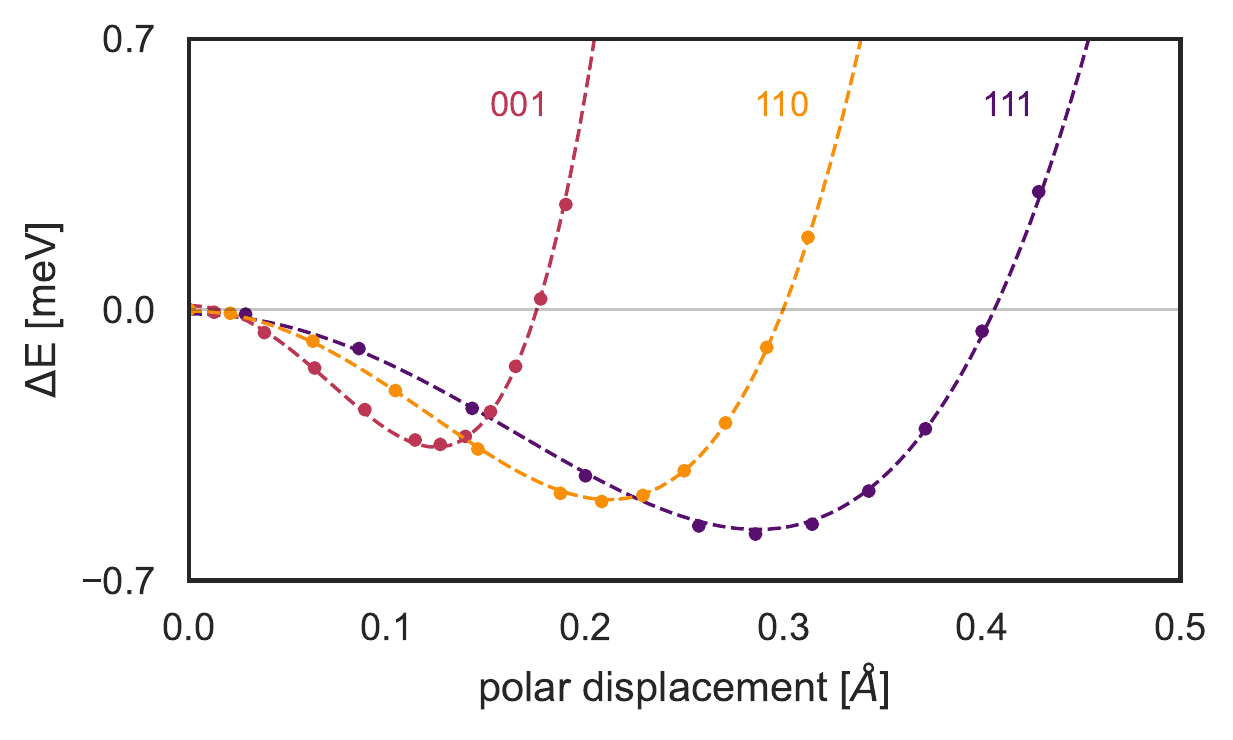}
	\caption{\label{fig:DW_directions}
        Calculated energy as a function of polar distortion along the high-symmetry directions in the doped (\SI{0.01}{e/fu}) KTO unit cell, as used in the bottom left of figure~\ref{fig:lambda_polar_soc} (`no SOC/polar').
        The horizontal axis shows the summed displacement of all atoms in a unit cell relative to their non-polar positions.
        A polarization along the [111] direction has both the largest energy gain and the largest displacement.
    }
\end{figure}

The fact that the polar symmetry breaking has such a large calculated effect on $\lambda$ is consistent with recent reports of polar symmetry breaking coexisting with superconductivity in STO \cite{Salmani-RezaiePolar2020,Salmani-RezaieRole2021,Salmani-RezaieRole2021} and theories of coupling to polar modes in KTO \cite{VendittiAnisotropic2022}.
Note that the tendency of KTO to become polar is strongest along the [111] direction, and weakest along the [001] direction, as shown in figure~\ref{fig:DW_directions}.
This trend is consistent with  that of the measured T\textsubscript{c}, again pointing to a possible relevance of the ferroelectric soft mode.

%-----------------------------------------------------------------------------
\section{Summary}
In summary, we have calculated the electron-phonon coupling in KTaO$_3$ for electron dopings between \SI{1.6e+18}{e/cm^3} and \SI{1.6e+21}{e/cm^3} and analyzed the results in light of the recently reported superconductivity and its surface dependence. 
Our calculations indicate that the measured trends in superconducting T$_c$ are not reflected in the calculated electron-phonon coupling strengths $\lambda$ along the corresponding reciprocal directions, confirming earlier suggestions that the superconductivity is not bulk BCS-like in nature \cite{RowleyFerroelectric2014,UenoFieldInduced2014,LiuTwodimensional2021,KleinTheory2022}. 
In this context, recent angle-resolved photoemission spectroscopy measurements implicating coupling of bulk-like electrons to Fuchs-Kliewer {\it surface} phonons are highly relevant \cite{ChenOrientationdependent2023}.
The concentration of $\lambda$ in the lowest frequency optical modes close to $\Gamma$ hints towards a mechanism in which the polar soft mode plays a role.
A comparison of the mode-resolved $\lambda$ values between non-polar and polar structures shows clearly that polarization strongly enhances the electron-phonon coupling.
In contrast, a comparison of mode-resolved $\lambda$ values between calculations without and with spin-orbit coupling shows negligible difference.

%-----------------------------------------------------------------------------
%-----------------------------------------------------------------------------
\section{Acknowledgments}\label{sec:acknowledgments}
The authors acknowledge helpful discussions with Jose Lorenzana, Manuel Bibes and Michael Fechner.
This work was funded by the European Research Council under the European Union’s Horizon 2020 Research and Innovation Program, grant agreement no.~810451. Computational resources were provided by ETH Zurich and the Swiss National Supercomputing Center (CSCS) under project ID s1128.

%-----------------------------------------------------------------------------
%-----------------------------------------------------------------------------
\section{Data availability}\label{sec:data}
All input data are available on the materialscloud archive at DOI:~\href{https://doi.org/10.24435/materialscloud:sn-xs}{10.24435/materialscloud:sn-xs} .

%-----------------------------------------------------------------------------
%-----------------------------------------------------------------------------
\newpage
\bibliography{Tobias_bibtex.bib}

%-----------------------------------------------------------------------------
%-----------------------------------------------------------------------------
\cleardoublepage
\newpage
\section*{Appendix}\label{sec:appendix}
\setcounter{figure}{0}
\renewcommand{\thefigure}{A\arabic{figure}}

This section contains data and information on the 2D to 3D charge carrier conversion, electronic band structures and numerical convergence of the calculations presented in the main text.

%-----------------------------------------------------------------------------
\subsection{Conversion between 2D and 3D carrier densities}
The interpolation used for conversion between $n_\mathrm{2D}$ and $n_\mathrm{3D}$ carrier densities is shown in figure~\ref{fig:interpol2D3D}. 
It is based on data from figure S6 of the supplementary information of Ref.~\onlinecite{UenoDiscovery2011}.

The corresponding interpolation formula is
\begin{equation*}
    n_\mathrm{3D} ~[\si{cm^{-3}}] = \SI{6.108e-3}{cm^{-1}} * n_\mathrm{2D}^{1.596} ~[\si{cm^{-2}}] \quad .    
\end{equation*}
According to this conversion, a 3D carrier concentration of \SI{1.4e20}{cm^{-3}} corresponds to \SI{1.0e+14}{cm^{-2}}, and \SI{2e21}{cm^{-3}} to \SI{5.4e+14}{cm^{-2}}.

\begin{figure}[H]
  \includegraphics[width=\linewidth, keepaspectratio]{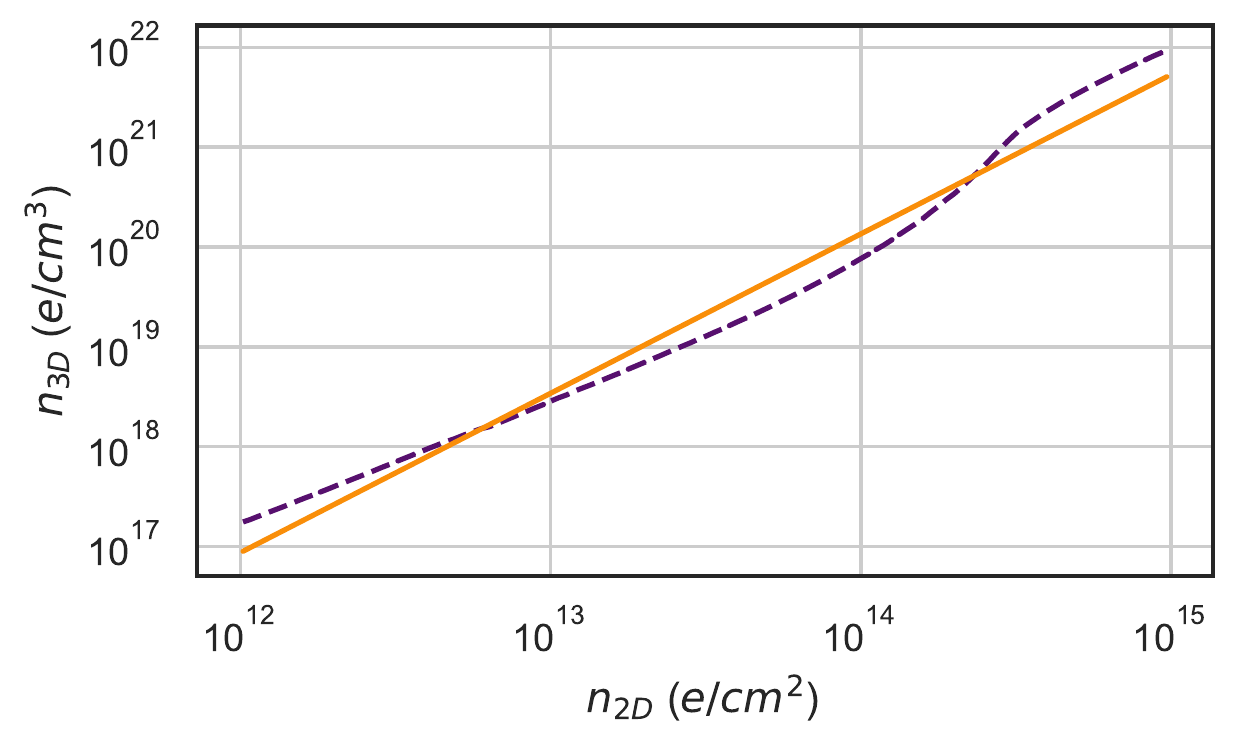}
  \caption{\label{fig:interpol2D3D}
    Interpolation used for conversion between $n_\mathrm{2D}$ and $n_\mathrm{3D}$ carrier densities (solid lighter line), based on data for KTO from figure S6 of the supplementary information of Ref.~\onlinecite{UenoDiscovery2011} (dashed dark line).
    The corresponding interpolation formula is $n_\mathrm{3D} = {6.1078*10^{-3}} * n_\mathrm{2D}^{1.5960}$.
  }
\end{figure}

%-----------------------------------------------------------------------------
\subsection{Convergence with coarse q mesh}
\begin{figure}[H]\vspace{1\baselineskip}
    \includegraphics[width=\linewidth,keepaspectratio]{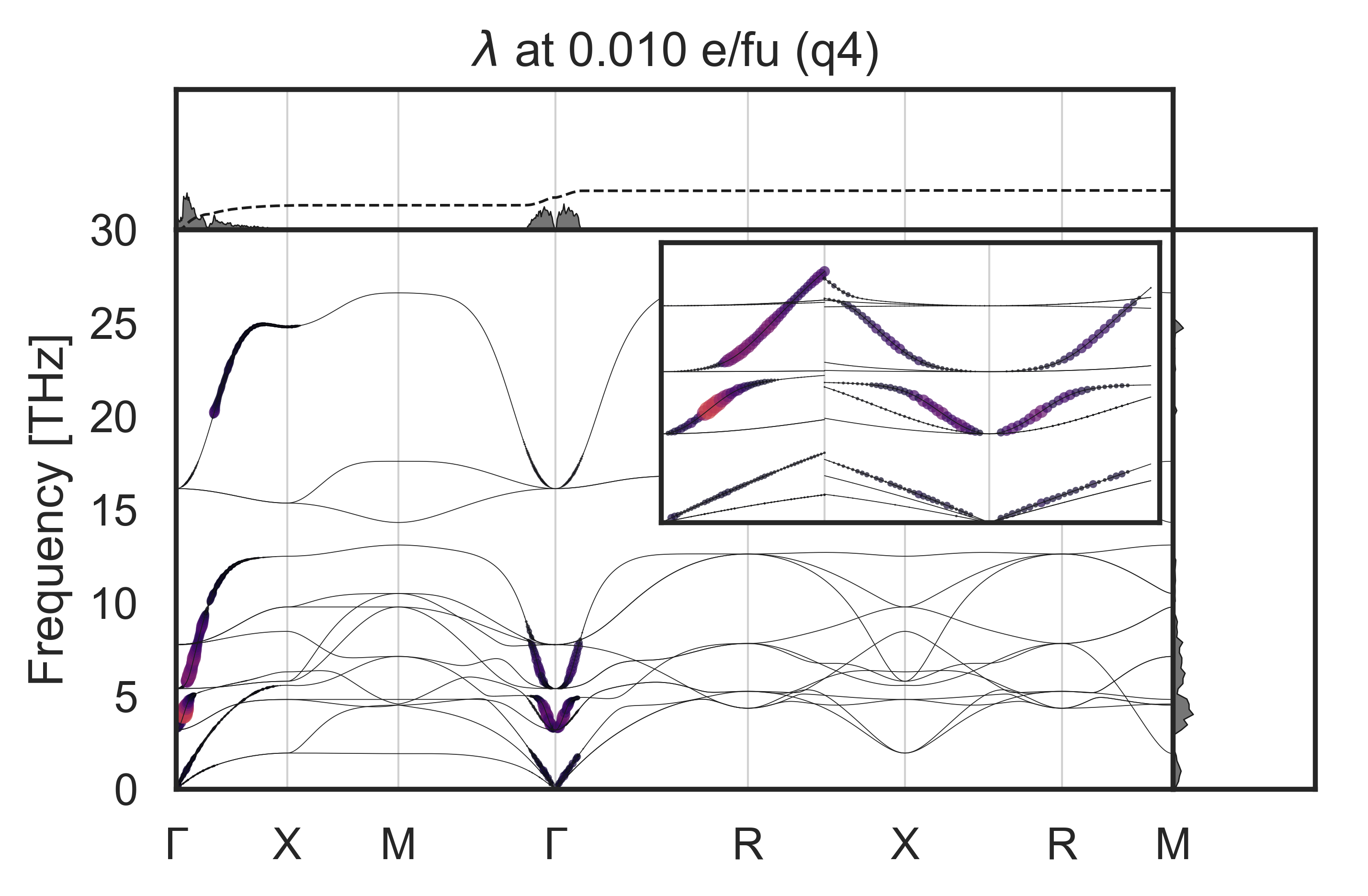}
    \includegraphics[width=\linewidth,keepaspectratio]{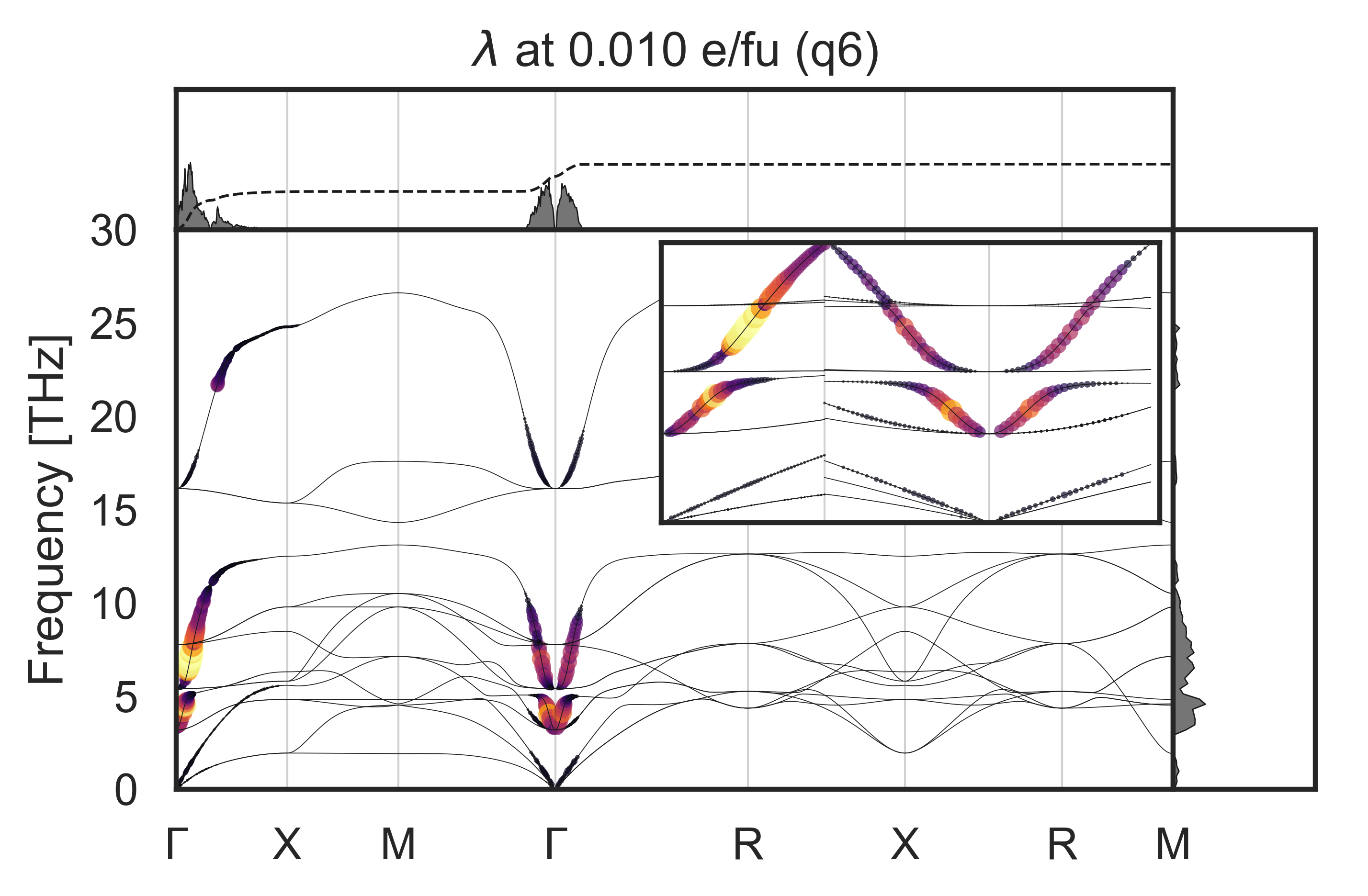}
    \includegraphics[width=\linewidth,keepaspectratio]{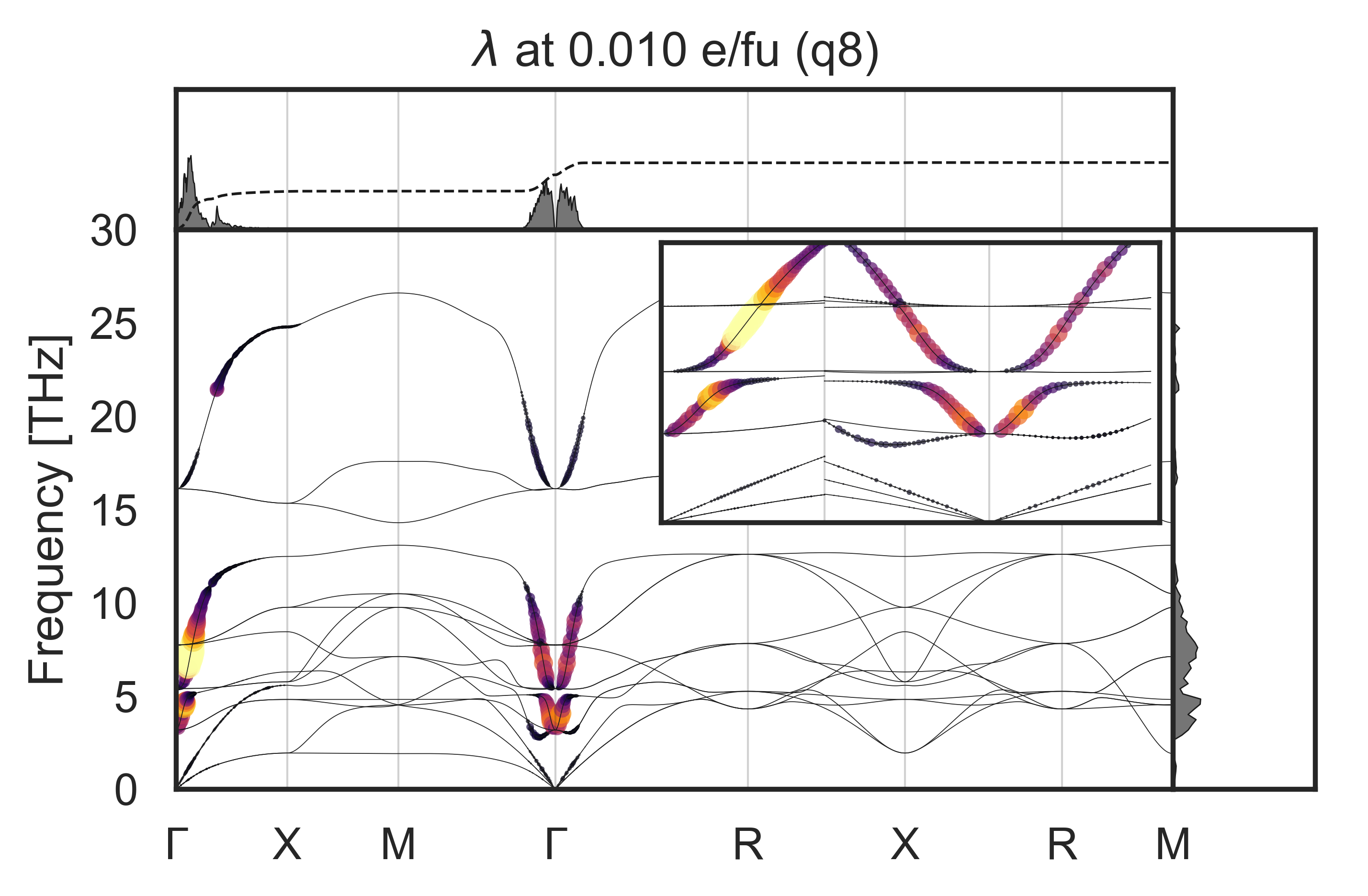}
    \caption{\label{fig:conv:lambda_qmesh}
       Calculated phonon dispersion and mode-resolved electron-phonon coupling strength $\lambda$ at a doping level of \SI{0.01}{e/fu} for q4, q6 and q8 meshes (top to bottom).
    }
\end{figure}

To test the convergence of our results, we calculate the phonons and the electron-phonon coupling strength $\lambda$ along the same high-symmetry path using three different q meshes (q4 ($4\times4\times4$), q6, and q8). Our results are shown in figure~\ref{fig:conv:lambda_qmesh}.
The phonon frequencies at the high-symmetry points are well converged at q4, with only minor variations between q4 and q6 or q8.
The colorscale is the same as that in fig.~\ref{fig:lambda_mode_resolved} of the main text.
The actual $\lambda$ values are about twice as large in the q6 and q8 cases compared to the q4 case.

%-----------------------------------------------------------------------------
\subsection{Decay properties in real space}
\begin{figure}[H]\vspace*{3mm}
    \includegraphics[width=0.5\linewidth, keepaspectratio]{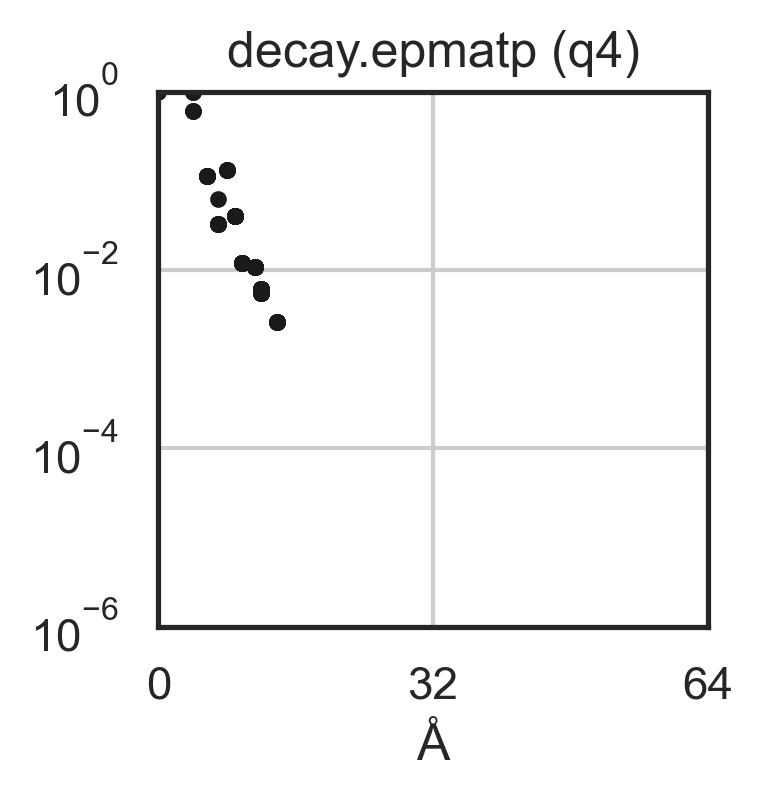}%
    \includegraphics[width=0.5\linewidth, keepaspectratio]{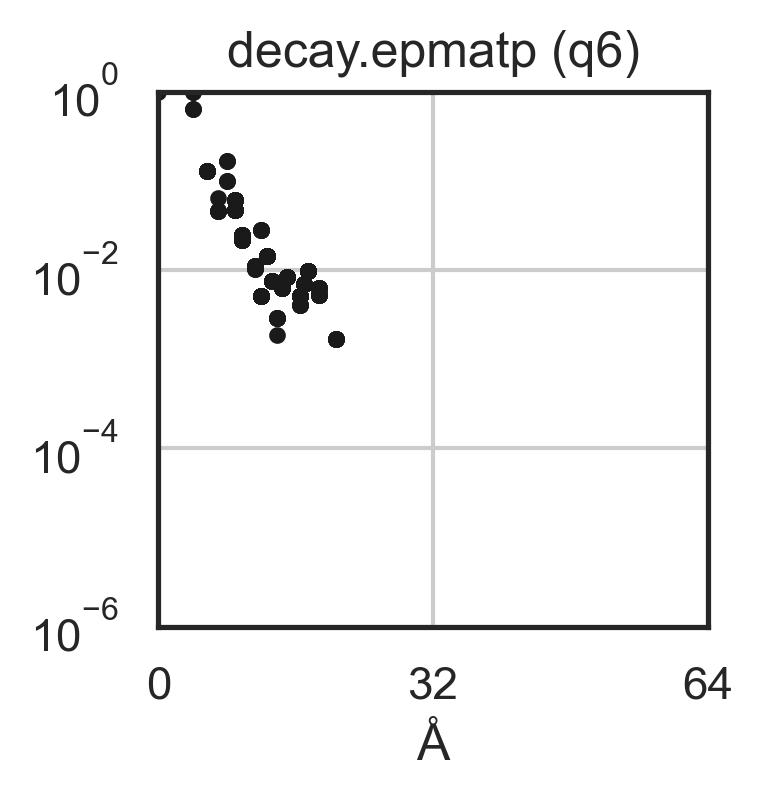}
    \includegraphics[width=0.5\linewidth, keepaspectratio]{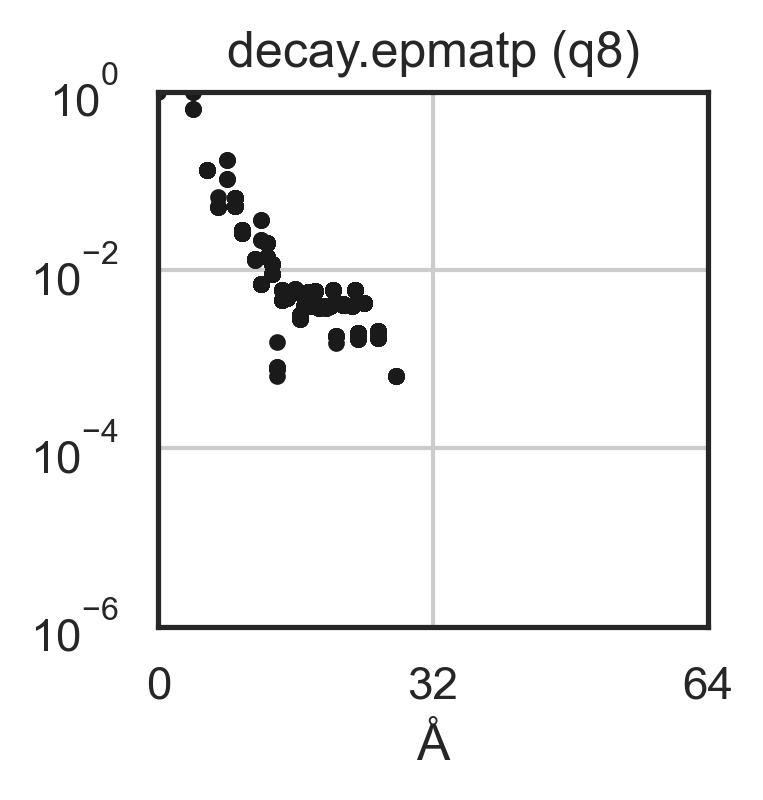}%
    \includegraphics[width=0.5\linewidth, keepaspectratio]{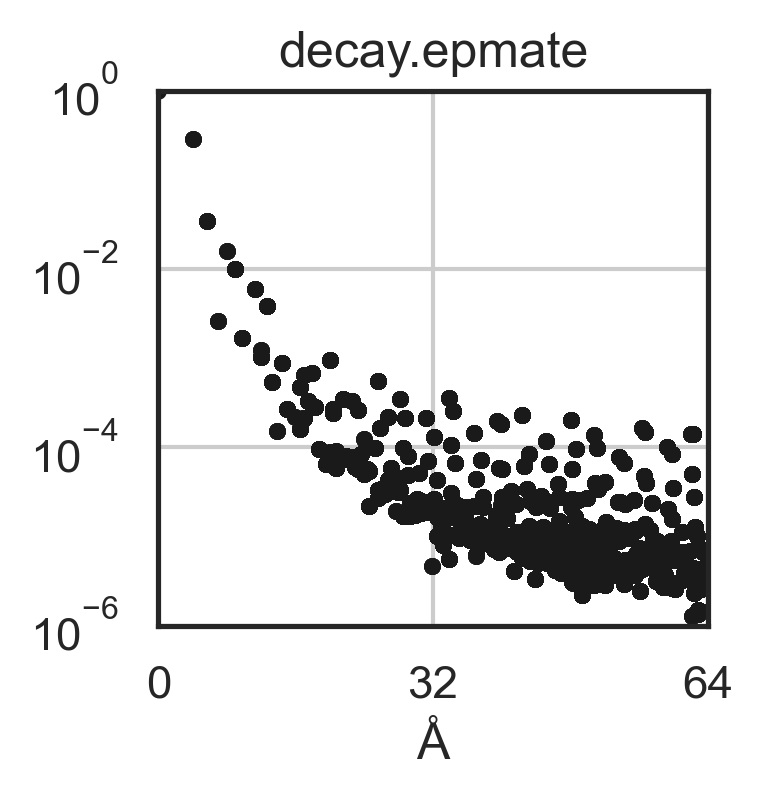}
    \caption{\label{fig:conv:decay}
        Decay of the phonon perturbation part of the electron-phonon matrix elements in real space for $4\times4\times4$ (q4), $6\times6\times6$ (q6) and $8\times8\times8$ (q8) q meshes (top left, top right and bottom left) and the electronic Wannier functions (bottom right), using a $24\times24\times24$ k mesh.
    }
\end{figure}

The spatial decay of the electron-phonon matrix elements in real space for the same three q meshes is shown in figure~\ref{fig:conv:decay} (see Refs.~\onlinecite{GiustinoElectronphonon2007} and \onlinecite{NoffsingerEPW2010} for more details).
We observe a decay of the phonon perturbation of almost three orders of magnitude using all three meshes.
The q6 and q8 meshes flatten after the decay, without further lowering the lowest value reached using the q4 mesh, indicating that the q4 mesh is accurate enough for the qualitative comparison we make.
The decay of the electronic Wannier functions is well converged using the $24\times24\times24$ coarse k-point mesh, as can be seen from the bottom right panel of figure~\ref{fig:conv:decay}.

%-----------------------------------------------------------------------------
\subsection{Electronic band structure with and without spin-orbit coupling}
\begin{figure}[H]
    \vspace*{2mm}
	\includegraphics[width=\linewidth,keepaspectratio]{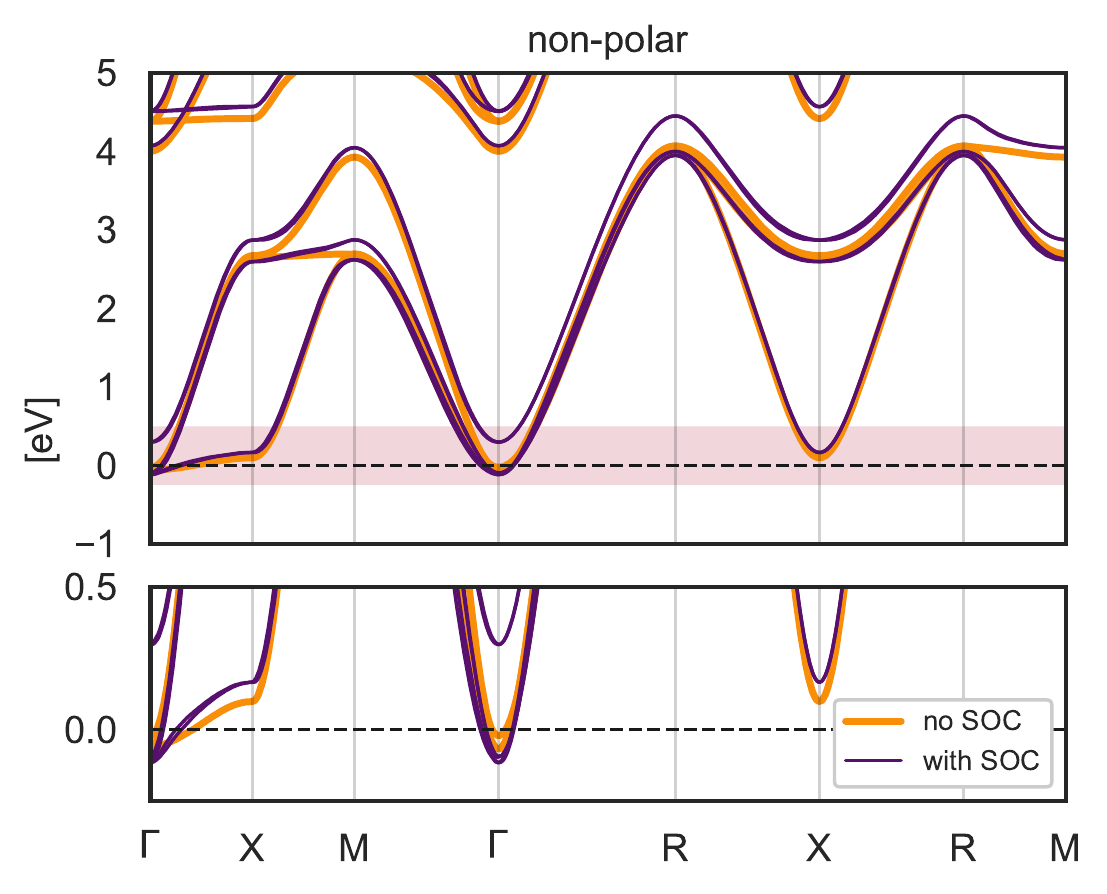}
	\includegraphics[width=\linewidth,keepaspectratio]{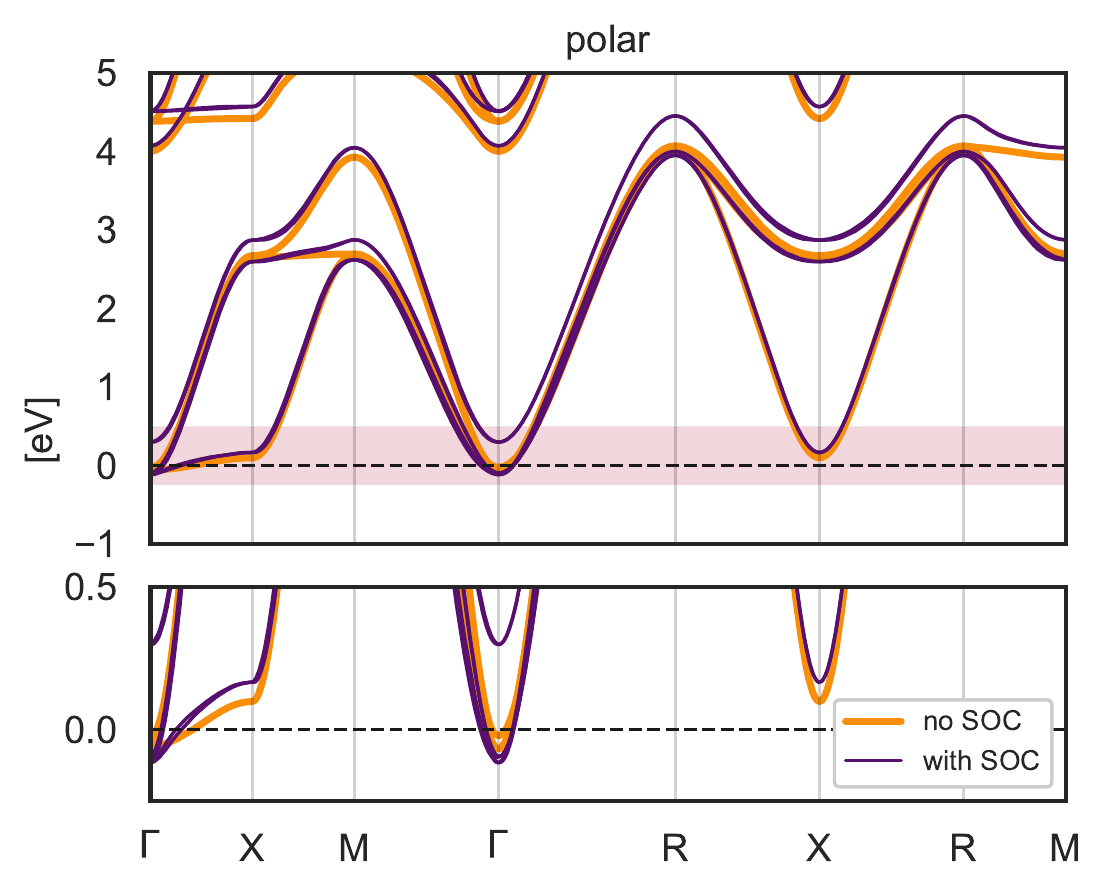}
	\caption{\label{fig:bands_SOC}
		Electronic bands of non-polar (upper plots) and polar (lower plots) \ce{KTaO3} without and with spin-orbit coupling (SOC), aligned at their respective Fermi energies (dashed lines) as calculated from the fine k mesh. 
        A band splitting of \SI{400}{meV} from SOC is observed in both cases.
        The polar band width is smaller due to the increases lattice constant necessary to stabilize the [111]-polar structure.
        The narrower bottom panel in each subplot is a zoom on the shaded region of each main panel
	}
\end{figure}

Figure~\ref{fig:bands_SOC} shows our calculated electronic band structure of non-polar (upper panel) and polar (lower panel) KTO without and with spin-orbit coupling, corresponding to figure~\ref{fig:lambda_polar_soc} in the main text.

\end{document}